%% file: main.tex
\documentclass{article}
\usepackage{PRIMEarxiv}

\input{mypackages}
\input{acronyms}

\title{ESBMC-PLC+: A Unified IEC~61131-3 Formal Verification Framework as a PLCverif Successor}

\author{
  Pierre Dantas \\
  Computer Science, The University of Manchester \\
  Manchester, UK \\
  \texttt{pierre.dantas@manchester.ac.uk} \\
  \And
  Lucas Cordeiro \\
  Computer Science, The University of Manchester \\
  Manchester, UK \\
  \texttt{lucas.cordeiro@manchester.ac.uk} \\
  \And
  Waldir Junior \\
  Electrical Engineering, UFAM \\
  Manaus, AM, Brazil \\
  \texttt{waldirjr@ufam.edu.br} \\
}

\begin{document}
\maketitle
\glsresetall

\begin{abstract}
PLCverif is the most mature open-source platform for \gls{plc} formal verification, developed at \gls{cern} and in production use since 2019. Yet it has two fundamental structural limitations: no support for \gls{ld} programs -- the dominant \gls{plc} notation -- and reliance on \gls{cbmc} as its primary backend, which restricts verification to bounded proofs. The PLCverif authors themselves identified \gls{esbmc} as the appropriate backend improvement. Prior work established ESBMC-PLC (native textual \gls{ld} frontend for \gls{esbmc} with \textit{k}-induction) and ESBMC-GraphPLC (graphical PLCopen XML support); together they cover \gls{ld} with unbounded proofs but not \gls{st} programs, and graphical \gls{ld} programs with timer or counter function blocks on rung paths remain unverifiable. This paper presents \textbf{ESBMC-PLC+}, a unified framework that closes the two remaining gaps: (1)~a \gls{st}/\gls{scl} frontend via the MATIEC open-source \glsfirst{iec}~61131-3 compiler, routing C-compiled \gls{st} programs to \gls{esbmc} with nondeterministic input modeling and YAML property injection; (2)~function block state semantics for graphical \gls{ld} -- extending the \gls{dfs} resolver to model \code{TON}/\code{TOF}/\code{TP} timers, \code{CTU}/\code{CTD} counters, and \code{R\_TRIG}/\code{F\_TRIG} edge triggers as persistent scan-cycle state variables in the GOTO~\gls{ir}. ESBMC-PLC+ is the first open-source \gls{plc} verification framework to accept all three major \glsfirst{iec}~61131-3 input formats (textual \gls{ld}, graphical \gls{ld}, and \gls{st}/\gls{scl}) through a single \gls{esbmc} backend providing \textit{k}-induction unbounded safety proofs. A systematic feature comparison with PLCverif and an experimental evaluation on 8~benchmark programs -- including programs with up to 8~integer-valued timer state variables -- demonstrate that ESBMC-PLC+ matches PLCverif's input coverage while providing strictly stronger verification guarantees. Against nuXmv's \gls{bdd} backend (PLCverif's unbounded prover), ESBMC-PLC+ is 400-2{,}000$\times$ faster on integer-timer programs and completes proofs where nuXmv \gls{bdd} times out at \SI{120}{\second}.
\end{abstract}


\glsresetall

\input{full_text}

\bibliographystyle{unsrtnat}
\bibliography{references}

\end{document}

%% file: mypackages.tex
\usepackage[utf8]{inputenc}
\usepackage[T1]{fontenc}
\usepackage{microtype}
\usepackage{nicefrac}

\usepackage{amsfonts}
\usepackage{amssymb}

\usepackage{xcolor}
\usepackage{colortbl}

\definecolor{amber}{rgb}{0.75, 0.50, 0.00}
\definecolor{teal} {rgb}{0.00, 0.50, 0.50}

\definecolor{reviewerblue} {RGB}{0,70,140}
\definecolor{commentgray}  {RGB}{240,240,245}
\definecolor{responsegreen}{RGB}{0,100,60}

\definecolor{grpFormal}{RGB}{180,160,220}
\definecolor{grpPlan}  {RGB}{140,210,185}
\definecolor{grpAgent} {RGB}{250,185,130}
\definecolor{grpRun}   {RGB}{250,215,100}
\definecolor{grpBench} {RGB}{195,195,185}

\definecolor{ldblue}  {HTML}{1565C0}
\definecolor{trgreen} {HTML}{1B5E20}
\definecolor{cpurp}   {HTML}{4A148C}
\definecolor{esorg}   {HTML}{BF360C}
\definecolor{safegrn} {HTML}{2E7D32}
\definecolor{unsafrd} {HTML}{B71C1C}
\definecolor{cexbrn}  {HTML}{4E342E}
\definecolor{diagpnk} {HTML}{880E4F}
\definecolor{panelbg} {HTML}{F5F5F5}

\definecolor{figviolet}{RGB}{180,160,220}
\definecolor{figamber} {RGB}{250,185,100}
\definecolor{figteal}  {RGB}{180,225,210}
\definecolor{figblue}  {RGB}{180,210,240}
\definecolor{figgray}  {RGB}{220,220,215}

\definecolor{codebg}    {HTML}{F8F8F8}
\definecolor{codeframe} {HTML}{DDDDDD}
\definecolor{codegreen} {HTML}{2E7D32}
\definecolor{codeblue}  {HTML}{1565C0}
\definecolor{codegray}  {HTML}{757575}
\definecolor{codepurple}{HTML}{7B1FA2}
\definecolor{codehl}    {gray}{0.95}

\colorlet{ldblue15}  {ldblue!15}
\colorlet{safegrn15} {safegrn!15}
\colorlet{esorg15}   {esorg!15}
\colorlet{trgreen15} {trgreen!15}
\colorlet{cpurp15}   {cpurp!15}
\colorlet{unsafrd15} {unsafrd!15}
\colorlet{cexbrn15}  {cexbrn!15}
\colorlet{diagpnk15} {diagpnk!15}

\usepackage{graphicx}
\usepackage{tikz}
\usetikzlibrary{
    positioning,
    shapes.geometric,
    shapes.misc,
    arrows.meta,
    calc,
    mindmap,
    shadows,
    fit,
    backgrounds,
    patterns
}
\usepackage{pgfplots}
\pgfplotsset{compat=1.18}

\usepackage{booktabs}
\usepackage{tabularx}
\usepackage{longtable}
\usepackage{multirow}
\usepackage{array}
\usepackage{makecell}
\usepackage{rotating}

\newcolumntype{L}[1]{>{\raggedright\arraybackslash}p{#1}}
\newcolumntype{C}[1]{>{\centering\arraybackslash}p{#1}}

\usepackage{listings}
\lstdefinelanguage{yaml}{
  keywords={true,false,null,yes,no},
  keywordstyle=\color{codeblue}\bfseries,
  comment=[l]{\#},
  commentstyle=\color{codegreen}\itshape,
  string=[b]",
  stringstyle=\color{codepurple},
  morestring=[b]{'},
  morestring=[b]{|},
  moredelim=[l][\color{amber}]{:},
  sensitive=true,
  breaklines=true,
  tabsize=2,
}

\usepackage{enumitem}
\usepackage{setspace}
\usepackage{comment}

\usepackage{siunitx}
\usepackage{eurosym}

\usepackage{algorithm}
\usepackage{algpseudocode}

\usepackage{soul}
\sethlcolor{codehl}
\newcommand{\code}[1]{\hl{\texttt{#1}}}

\usepackage{url}
\usepackage{hyperref}
\hypersetup{
  colorlinks = true,
  linkcolor  = blue,
  citecolor  = blue,
  filecolor  = blue,
  urlcolor   = blue,
  pdfborder  = {0 0 0},
}
\usepackage[numbers]{natbib}

\usepackage[acronym, nogroupskip, nonumberlist, nopostdot]{glossaries}
\glsdisablehyper

\newcommand{\bars}[1]{%
  \begin{tikzpicture}[baseline=-0.4ex]
    \foreach \i in {1,...,5}{%
      \pgfmathparse{\i <= #1 ? 1 : 0}%
      \ifnum\pgfmathresult=1
        \fill[black]     (\i*0.12-0.12, 0) rectangle (\i*0.12-0.03, 0.22);
      \else
        \fill[black!18]  (\i*0.12-0.12, 0) rectangle (\i*0.12-0.03, 0.22);
      \fi
    }
  \end{tikzpicture}%
}

%% file: acronyms.tex
\newacronym{aadl}{AADL}{Architecture Analysis and Design Language}
\newacronym{ansi-c}{ANSI-C}{American National Standards Institute C}
\newacronym{api}{API}{Application Programming Interface}
\newacronym{ast}{AST}{Abstract Syntax Tree}
\newacronym{bdd}{BDD}{Binary Decision Diagrams}
\newacronym{bmc}{BMC}{Bounded Model Checking}
\newacronym{cbmc}{CBMC}{Bounded Model Checking for ANSI-C Programs}
\newacronym{cegar}{CEGAR}{Counterexample-Guided Abstraction Refinement}
\newacronym{cern}{CERN}{Conseil Européen pour la Recherche Nucléaire}
\newacronym{cfg}{CFG}{Control Flow Graph}
\newacronym{chc}{CHC}{Constrained Horn Clause}
\newacronym{cli}{CLI}{Command-Line Interface}
\newacronym{cpu}{CPU}{Central Processing Unit}
\newacronym{ctl}{CTL}{Computation Tree Logic}
\newacronym{cuda}{CUDA}{Compute Unified Device Architecture}
\newacronym{cve}{CVE}{Common Vulnerability and Exposure}
\newacronym{dfs}{DFS}{Depth-First Search}
\newacronym{dsl}{DSL}{Domain-Specific Language}
\newacronym{epsrc}{EPSRC}{Engineering and Physical Sciences Research Council}
\newacronym{esbmc}{ESBMC}{Efficient SMT-based Context-Bounded Model Checker}
\newacronym{fbd}{FBD}{Functional Block Diagram}
\newacronym{fpga}{FPGA}{Field-Programmable Gate Array}
\newacronym{ic3}{IC3}{Incremental Construction of Inductive Clauses for Indubitable Correctness}
\newacronym{ide}{IDE}{Integrated Development Environment}
\newacronym{iec}{IEC}{International Electrotechnical Commission}
\newacronym{ieee}{IEEE}{Institute of Electrical and Electronics Engineers}
\newacronym{ics}{ICS}{Industrial Control Systems}
\newacronym{il}{IL}{Instruction List}
\newacronym{iot}{IoT}{Internet of Things}
\newacronym{ir}{IR}{Intermediate Representation}
\newacronym{iso}{ISO}{International Organization for Standardization}
\newacronym{ld}{LD}{Ladder Diagram}
\newacronym{llb}{LLB}{Ladder Logic Bombs}
\newacronym{llm}{LLM}{Large Language Model}
\newacronym{ltl}{LTL}{Linear Temporal Logic}
\newacronym{nasa}{NASA}{National Aeronautics and Space Administration}
\newacronym{pdr}{PDR}{Property Directed Reachability}
\newacronym{pii}{PII}{Process Image Input}
\newacronym{pio}{PIO}{Process Image Output}
\newacronym{plc}{PLC}{Programmable Logic Controller}
\newacronym{por}{POR}{Partial Order Reduction}
\newacronym{pou}{POU}{Program Organization Unit}
\newacronym{rtos}{RTOS}{Real-Time Operating System}
\newacronym{sat}{SAT}{Boolean Satisfiability}
\newacronym{scl}{SCL}{Structured Control Language}
\newacronym{sfc}{SFC}{Sequential Function Chart}
\newacronym{slr}{SLR}{Systematic Literature Review}
\newacronym{smt}{SMT}{Satisfiability Modulo Theories}
\newacronym{smtlib2}{SMT-LIB}{Satisfiability Modulo Theories Library}
\newacronym{smv}{SMV}{Symbolic Model Verifier}
\newacronym{ssa}{SSA}{Static Single Assignment}
\newacronym{st}{ST}{Structured Text}
\newacronym{stl}{STL}{Statement List}
\newacronym{svcomp}{SV-COMP}{Competition on Software Verification}
\newacronym{tacas}{TACAS}{Tools and Algorithms for the Construction and Analysis of Systems}
\newacronym{ufam}{UFAM}{Federal University of Amazonas}
\newacronym{ukri}{UKRI}{UK Research and Innovation}

%% file: full_text.tex
\section{Introduction}
\label{sec:intro}

\glspl{plc} execute safety-critical programs across nuclear power stations, water treatment plants, chemical refineries, and railway signaling systems. Formal verification -- mathematical proof that a program satisfies a specification for all executions -- is the appropriate assurance technique for such deployments. Yet, the gap between formal verification research and \gls{plc} engineering practice remains wide. Two barriers are responsible: the \emph{language barrier} (formal tools historically target C or \gls{smv}, not \gls{iec}~61131-3 languages) and the \emph{backend barrier} (\gls{plc}-oriented tools typically use \glspl{bmc} that cannot produce unbounded safety proofs).

\textbf{PLCverif}~\cite{LopezMiguel2022, LopezMiguel2025}, developed at \gls{cern} and in production use since 2019, is the most mature open-source response to this challenge. It accepts Siemens \gls{scl} (a \gls{st} dialect) and routes programs to \gls{cbmc}, nuXmv, or Theta. PLCverif has been deployed industrially for the GSI heavy-ion accelerator~\cite{LopezMiguel2025} and extended to support pure-past \gls{ltl} safety properties via FRET~\cite{Fink2024}. Yet PLCverif has three structural limitations that bound its applicability:

\begin{enumerate}[leftmargin=2em]
  \item \textbf{No \gls{ld} input.} PLCverif requires \gls{scl} or \gls{stl}. \gls{ld} programs -- the most widely deployed \gls{plc} notation, accounting for over 60\% of installed bases in North American and Japanese manufacturing~\cite{Weiss2021} -- must be manually translated, introducing fidelity risks and significant engineer effort for large programs.
  
  \item \textbf{Bounded backend.} PLCverif's primary backend, \gls{cbmc}, is a \gls{bmc} that cannot prove safety for all future scan cycles. Its nuXmv backend provides \gls{bdd}-based unbounded proofs that scale poorly to programs with large integer variables. The PLCverif authors themselves identified \gls{esbmc} as the appropriate direction for improvement, noting that \emph{``an \gls{smt}-based model checker like \gls{esbmc} could improve the performance of \gls{cbmc}''}~\cite{LopezMiguel2022}.
  
  \item \textbf{Incomplete graphical \gls{ld} coverage.} Even when \gls{ld}$\to$\gls{st} conversion is attempted, graphical PLCopen XML programs containing timer or counter function blocks on rung paths are not correctly handled by existing open-source tools.
\end{enumerate}

\noindent Prior work addressed the first limitation from the \gls{esbmc} direction. \textbf{ESBMC-PLC}~\cite{DantasCordeiro2026artefact} introduced the first open-source formal verifier with native support for textual PLCopen XML \gls{ld}, implementing an \gls{ld} frontend for \gls{esbmc} and providing unbounded safety proofs via \textit{k}-induction. \textbf{ESBMC-GraphPLC}~\cite{DantasCordeiro2026graphical} extended ESBMC-PLC to graphical PLCopen XML using a \gls{dfs}-based rung extractor, but explicitly excluded programs whose rung paths contain function blocks (timers, counters, edge triggers), reporting this as the highest-priority future work.

\begin{figure}[htbp]
  \centering
  \includegraphics[width=0.75\linewidth]{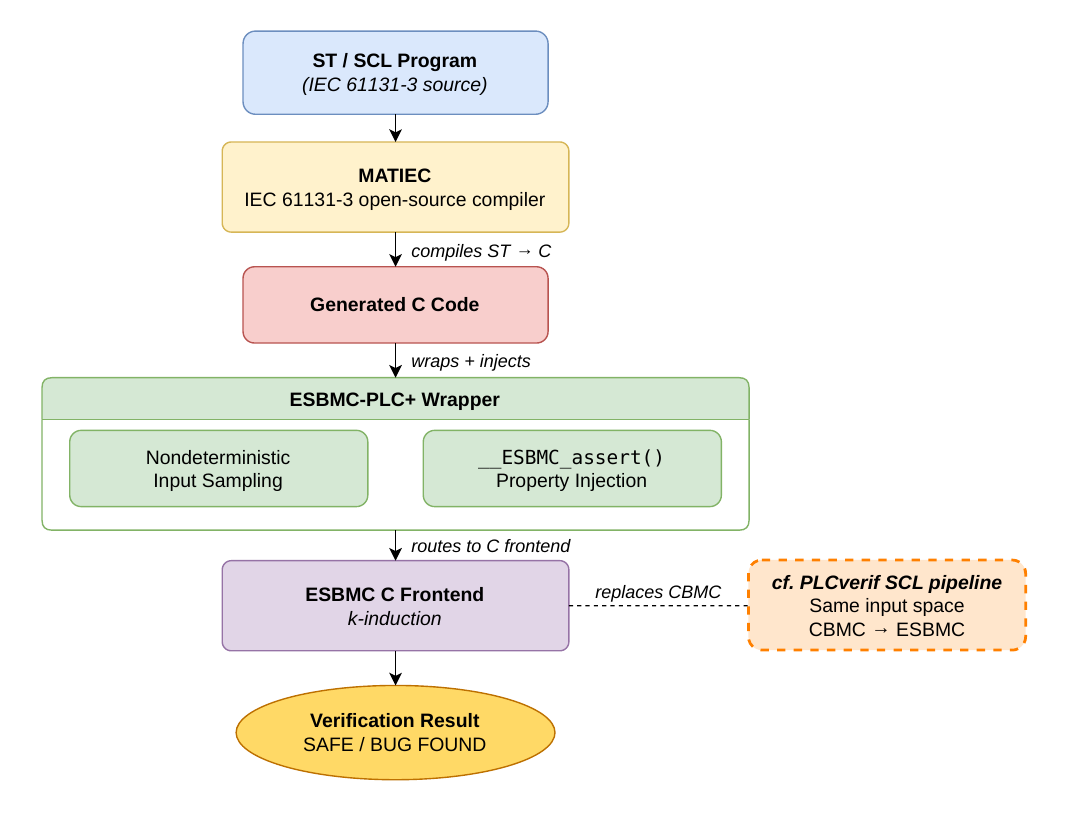}
  \caption{ESBMC-PLC+ \gls{st}/\gls{stl} verification pipeline via MATIEC: the compiler translates \gls{iec}~61131-3 \gls{st}/\gls{stl} to C, which is wrapped with nondeterministic input sampling and \code{\_\_ESBMC\_assert()} property injection before being routed to ESBMC's C frontend with \textit{k}-induction, replacing \gls{cbmc} in the PLCverif \gls{scl} pipeline.}
  \label{fig:matiec-pipeline}
\end{figure}

This paper presents \textbf{ESBMC-PLC+}~\cite{ESBMCpr5400}, a unified framework that completes the path to a PLCverif successor by closing the two remaining gaps:

\begin{enumerate}[leftmargin=2em]
    \item \textbf{\gls{st}/\gls{scl} frontend via MATIEC} (\S\ref{sec:st-frontend}): ESBMC-PLC+ integrates the MATIEC open-source \gls{iec}~61131-3 compiler~\cite{deSousa2014} to accept \gls{st} programs. MATIEC compiles \gls{st} to C; ESBMC-PLC+ wraps the generated C with nondeterministic input sampling and \code{\_\_ESBMC\_assert()} property injection, routing it to \gls{esbmc}'s C frontend with \textit{k}-induction (Figure~\ref{fig:matiec-pipeline}). This covers the same input space as PLCverif's \gls{scl} pipeline while replacing \gls{cbmc} with \gls{esbmc}.

    \item \textbf{Graphical \gls{ld} function block semantics} (\S\ref{sec:fb-extension}): The \gls{dfs} resolver from ESBMC-GraphPLC is extended to model \code{TON}/\code{TOF}/\code{TP} timers, \code{CTU}/\code{CTD} counters, and \code{R\_TRIG}/\code{F\_TRIG} edge triggers as persistent scan-cycle state variables in the GOTO~\gls{ir}. This closes the gap that excluded \code{beremiz\_traffic\_light} and similar timer-gated programs from ESBMC-GraphPLC's evaluation.
\end{enumerate}

Together with the inherited ESBMC-PLC and ESBMC-GraphPLC contributions, ESBMC-PLC+ becomes the first open-source verification framework to accept all three major \gls{iec}~61131-3 input formats -- textual \gls{ld}, graphical \gls{ld}, and \gls{st}/\gls{scl} -- through a single \gls{esbmc} backend providing \textit{k}-induction unbounded safety proofs.

\subsection{Contributions}

\begin{enumerate}[leftmargin=2em]
  \item \textbf{Unified ESBMC-PLC+ architecture} (\S\ref{sec:architecture}): a single framework integrating three input frontends (textual \gls{ld}, graphical \gls{ld} with function blocks, \gls{st}/\gls{scl} via MATIEC) under a shared \gls{esbmc} backend and unified \gls{cli}.

  \item \textbf{\gls{st}/\gls{scl} frontend via MATIEC} (\S\ref{sec:st-frontend}): a translation pipeline from \gls{iec}~61131-3 \gls{st} programs to \gls{esbmc} via MATIEC-generated C, with nondeterministic input modeling and YAML property injection matching the ESBMC-PLC property language.

  \item \textbf{Graphical \gls{ld} function block state model} (\S\ref{sec:fb-extension}): scan-cycle-accurate GOTO~\gls{ir} encodings for six function block types (\code{TON}, \code{TOF}, \code{TP}, \code{CTU}, \code{CTD}, \code{R\_TRIG}/\code{F\_TRIG}), grounded in Ebnenasir's formal \gls{ld} semantics~\cite{Ebnenasir2023} and confined to the single source file changed by ESBMC-GraphPLC.

  \item \textbf{Experimental evaluation and PLCverif comparison} (\S\ref{sec:experiments}--\ref{sec:plcverif-compare}): evaluation on 18~benchmark programs (13~inherited ESBMC-PLC textual \gls{ld}, 3~inherited ESBMC-GraphPLC graphical \gls{ld}, 1~new graphical \gls{ld} with function blocks, 1~new \gls{st} program) demonstrating correct classification with zero false positives and zero regressions on inherited benchmarks. A direct nuXmv~\gls{bdd} vs.\ ESBMC-PLC+ comparison on 8~\gls{ld} benchmark runs (4~programs $\times$ safe/unsafe variants) shows ESBMC-PLC+ is 400--2{,}000$\times$ faster on timer-intensive programs and completes proofs that nuXmv \gls{bdd} cannot finish within a \SI{120}{\second} timeout.
\end{enumerate}

\subsection{Scope}

ESBMC-PLC+ inherits the scope of ESBMC-PLC and ESBMC-GraphPLC for \gls{ld} programs: contacts, coils (\code{XIC}/\code{XIO}/\code{OTE}/\code{OTL}/\code{OTU}), timers (\code{TON}/\code{TOF}/\code{TP}), counters (\code{CTU}/\code{CTD}), and arithmetic function blocks; REAL/FLOAT types, strings, arrays, and multi-\gls{pou} programs remain unsupported for \gls{ld}. The \gls{st} frontend inherits MATIEC's coverage, which includes the full \gls{iec}~61131-3 \gls{st} core, including REAL types, arrays, and multi-\gls{pou}. Function block handling in graphical \gls{ld} covers the six types listed above; vendor-specific function blocks generate an \code{UnsupportedFB} warning and fall back to the nondeterministic over-approximation from ESBMC-GraphPLC.


This paper is structured as a progression from problem identification to solution design, implementation, and empirical validation. Sections~\ref{sec:background}--\ref{sec:plcverif-analysis} provide background and analyse PLCverif's limitations; Sections~\ref{sec:architecture}--\ref{sec:fb-extension} describe the architecture and the two main contributions (\S\ref{sec:st-frontend} and \S\ref{sec:fb-extension}); Sections~\ref{sec:experiments}--\ref{sec:plcverif-compare} present the experimental evaluation and comparison with PLCverif (key results in Tables~\ref{tab:benchmarks}, \ref{tab:feature-compare}, \ref{tab:nuxmv_comparison}, and \ref{tab:nuxmv_comparison_normalized}); Sections~\ref{sec:discussion}--\ref{sec:conclusion} discuss limitations, future work, and conclude.

\section{Background}
\label{sec:background}

\subsection{\gls{plc} Architecture and the Scan Cycle}

\begin{figure}[htbp]
\centering
  \includegraphics[width=1\linewidth]{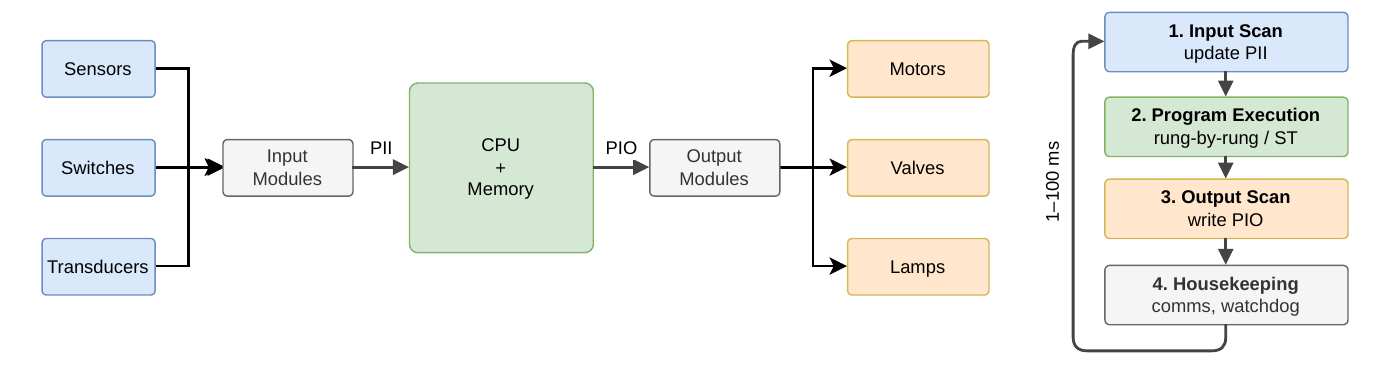}
    \caption{Left: PLC hardware architecture. Sensor signals are latched into the \gls{pii}; the \gls{cpu} executes the user program; results drive actuators via the process image output (\gls{pio}). Right: the scan cycle repeats indefinitely; all four steps are complete within one cycle period.}
\label{fig:plc-arch-cycle}
\end{figure}

A \gls{plc} is an industrial digital computer engineered for deterministic, real-time control of manufacturing and process automation systems. As illustrated in Figure~\ref{fig:plc-arch-cycle}-left, a \gls{plc} comprises four principal hardware components: a \gls{cpu} that executes the user control program; non-volatile \emph{program memory} that retains both the program and retentive variable values across power cycles; \emph{input modules} that digitise and latch physical sensor signals (pushbuttons, limit switches, thermocouples, pressure transducers) into a \gls{pii} at the start of each cycle; and \emph{output modules} that drive physical actuators (solenoid valves, motor contactors, indicator lamps) from a \gls{pio} written at cycle end.

Execution follows a strictly deterministic, cyclic model called the \emph{scan cycle} (Figure~\ref{fig:plc-arch-cycle}-right):
\begin{enumerate}[leftmargin=2em]
  \item \textbf{Input scan.} All input module registers are read and copied atomically into the \gls{pii}, freezing a consistent snapshot of the physical world for the entire scan.
  \item \textbf{Program execution.} The user program executes sequentially -- rung by rung in \gls{ld}, or statement by statement in \gls{st} -- reading exclusively from the \gls{pii} and writing results to the \gls{pio} and internal state variables.
  \item \textbf{Output scan.} The completed \gls{pio} is copied to the output modules, which drive the physical actuators.
  \item \textbf{Housekeeping.} The \gls{cpu} services fieldbus and Ethernet communications, updates diagnostic registers, and resets the watchdog timer to confirm a healthy scan.
  \item \textbf{Repeat.}
\end{enumerate}
Cycle times typically range from \SIrange{1}{100}{\milli\second}, determined by program size, I/O count, and \gls{cpu} clock speed.

The cyclic execution model has a direct implication for formal verification: because the \gls{pii} snapshot is taken \emph{once} per cycle and held constant during program execution, a \gls{plc} program behaves as a \emph{synchronous reactive system} -- at each cycle boundary, nondeterministic inputs arrive, the program function is applied, and new outputs are produced. ESBMC-PLC+ encodes this as a \code{while(true)} loop in which input variables are re-sampled via \code{nondet\_bool()} and \code{nondet\_int()} (open-world sensor model), output coils and timer/counter state are persistent \code{static} variables (process-image semantics), and user-defined safety properties are injected as \code{\_\_ESBMC\_assert()} statements inside the loop body. The \textit{k}-induction engine then proves that all assertions hold for every reachable state across all future scan counts -- an unbounded safety proof over the infinite scan-cycle trajectory.

\subsection{\mbox{\gls{iec}~61131-3} Languages}

Table~\ref{tab:iec-languages} summarises the five \mbox{\gls{iec}~61131-3} languages and the support each receives in PLCverif and ESBMC-PLC+. \gls{ld} dominates installed base, particularly in North American and Japanese manufacturing. A Ladder program consists of \emph{rungs}: each rung evaluates a Boolean combination of \emph{contacts} and assigns the result to one or more \emph{coils}. Timer and counter function blocks appear as specialized rung elements. \gls{st} is syntactically similar to Pascal and is the natural target for source-to-source translation to C; it is used for complex algorithms and is the input accepted by PLCverif's \gls{scl} pipeline.

\begin{table}[htbp]
\centering
\small
\caption{\mbox{\gls{iec}~61131-3} programming languages and tool support}
\label{tab:iec-languages}
\begin{tabular}{llll}
\toprule
\textbf{Language} & \textbf{Type} & \textbf{PLCverif} & \textbf{ESBMC-PLC+} \\
\midrule
\acrfull{ld} (graphical)       & Graphical & No (manual) & \textbf{Yes (native)} \\
\gls{fbd}                       & Graphical & Via OpennessScripter & No \\
\gls{st} / \gls{scl}           & Textual   & Yes (\gls{scl}) & \textbf{Yes (MATIEC)} \\
\gls{il} (deprecated)          & Textual   & No & No \\
\gls{sfc}                       & Graphical & No & No \\
\bottomrule
\end{tabular}
\end{table}

\subsection{\mbox{\gls{esbmc}}: Verification Engine}

\begin{figure}[htbp]
\centering
    \includegraphics[width=1\linewidth]{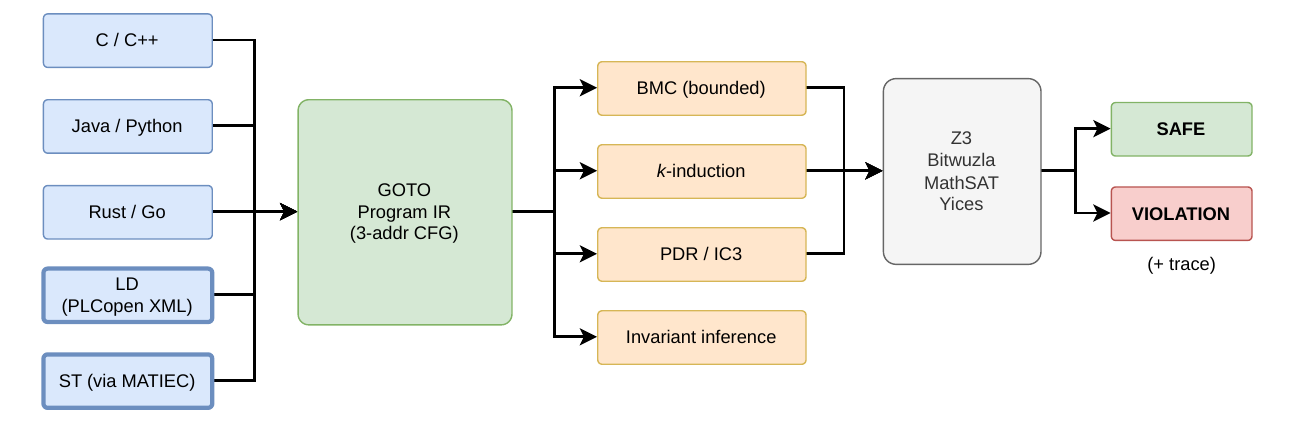}
    \caption{\gls{esbmc} internal architecture. Multiple source-language frontends compile to a unified GOTO program \gls{ir}; four verification algorithms consume the \gls{ir} and delegate queries to pluggable \gls{smt} solvers. ESBMC-PLC+ contributes the two bold-bordered frontends (\gls{ld} and \gls{st}); the GOTO \gls{ir} and the backend are unchanged.}
\label{fig:esbmc-arch}
\end{figure}

\gls{esbmc}~\cite{menezes2024, gadelha2020} is an open-source, multi-language, \gls{smt}-based model checker developed at the University of Manchester, the University of Southampton, and partner institutions. Its internal architecture, shown in Figure~\ref{fig:esbmc-arch}, is organized around a single, language-independent intermediate representation: the \emph{GOTO program}.

\textbf{GOTO intermediate representation.} Every \gls{esbmc} frontend compiles its source language into a GOTO program -- a low-level, three-address control-flow graph with explicit \code{GOTO}, \code{ASSUME}, and \code{ASSERT} instructions, typed variables, and structured function calls. Once a program is in GOTO form, the entire verification backend applies uniformly regardless of the source language. This is precisely why ESBMC-PLC+ can introduce new frontends (the \gls{ld} parsers, the MATIEC-based \gls{st} wrapper) without modifying the backend: they produce GOTO programs that the existing engine already processes.

\textbf{Verification algorithms.} \gls{esbmc} supports four complementary proof strategies:
\begin{enumerate}[leftmargin=2em]
  \item \textbf{\gls{bmc}.} The scan-cycle loop is unrolled to bound $k$, yielding a loop-free formula that is discharged by an \gls{smt} solver. If satisfiable, a concrete counterexample trace is extracted. If unsatisfiable, the property holds for all executions of length $\leq k$ -- a bounded certificate.
  \item \textbf{\textit{k}-induction.} The \emph{base case} establishes that no violation occurs in the first $k$ steps; the \emph{inductive step} proves that if the property holds across any window of $k$ consecutive steps, it must hold in the next step. Together, these two sub-goals constitute an \emph{unbounded} safety proof -- valid for \emph{all} scan counts, not just a finite prefix. This is the algorithm used by ESBMC-PLC+ for all three frontends.
  \item \textbf{\gls{pdr} / \gls{ic3}.} A frame-based algorithm that incrementally refines inductive invariants; effective for purely Boolean state spaces and programs with shallow proof depth.
  \item \textbf{Invariant inference.} Abstract interpretation automatically synthesizes loop invariants that can strengthen the inductive step of \textit{k}-induction, reducing the required bound $k$.
\end{enumerate}

\textbf{\gls{smt} backends.} All satisfiability queries are delegated to a pluggable \gls{smt} solver. Supported solvers include Z3~\cite{demoura2008z3}, Bitwuzla, MathSAT, Yices, and Boolector. For \gls{plc} programs with integer timer and counter state, the bit-vector arithmetic theory (\code{QF\_BV}) is critical: it allows \gls{esbmc} to reason about all $2^{16}=65{,}536$ values of a 16-bit accumulator in a \emph{single symbolic query}, rather than enumerating states as a \gls{bdd}-based prover must.

\subsection{ESBMC-PLC and ESBMC-GraphPLC}

\textbf{ESBMC-PLC}~\cite{DantasCordeiro2026artefact} implements an \gls{ld} frontend for \gls{esbmc} that parses textual PLCopen XML, translates \gls{ld} rungs to GOTO~\gls{ir}, and checks user-defined safety properties via \gls{bmc} or \textit{k}-induction. Its five-kind YAML property language (\code{mutual\_exclusion}, \code{invariant}, \code{absence}, \code{response}, \code{reachability}) enables specification without temporal logic expertise. The experimental evaluation on 13 benchmarks spanning 6~industrial domains demonstrated correct classification of all programs, 8~bugs found, and all 61~properties checked in under \SI{60}{\milli\second}.

\textbf{ESBMC-GraphPLC}~\cite{DantasCordeiro2026graphical} extends ESBMC-PLC to graphical PLCopen XML via a \gls{dfs}-based resolver that traverses the \code{localId}/\code{refLocalId} connection graph. The extension is confined to 274~lines in \code{plcopen\_xml\_parser.cpp}; no backend change was required. The resolver drops function block nodes from rung expressions and reports timer-gated programs (e.g., \code{beremiz\_traffic\_light}) as an explicitly documented limitation that requires future work.

\subsection{PLCverif}

PLCverif~\cite{LopezMiguel2022, LopezMiguel2025} is a formal verification framework developed at \gls{cern} and deployed in production since 2019 for safety-critical control programs of the \gls{cern} SPS-PPS accelerator complex and the GSI heavy-ion accelerator~\cite{LopezMiguel2025}. It is the most mature open-source \gls{plc} formal verification tool and the closest prior work to ESBMC-PLC+.

\begin{figure}[htbp]
\centering
    \includegraphics[width=1\linewidth]{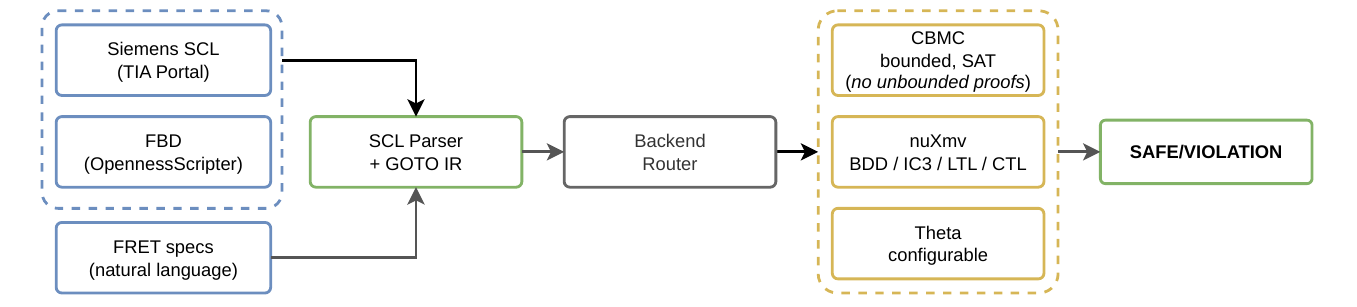}
    \caption{PLCverif architecture. Siemens \gls{scl} and \gls{fbd} inputs are parsed into a GOTO-compatible \gls{ir}; FRET specifications are compiled to nuXmv \gls{ltl} patterns. The backend router dispatches to \gls{cbmc} (bounded only), nuXmv (unbounded, full temporal logic), or Theta.}
\label{fig:plcverif-arch}
\end{figure}

\textbf{Architecture.} The PLCverif pipeline, shown in Figure~\ref{fig:plcverif-arch}, accepts programs written in Siemens \gls{scl} -- a proprietary \gls{st} dialect used in TIA Portal -- or, via the OpennessScripter bridge, \gls{fbd} exported from TIA Portal. The frontend parses the input and produces a GOTO-like \gls{ir} compatible with \gls{cbmc}'s internal format. This \gls{ir} is then routed to one of three backends:

\begin{enumerate}[leftmargin=2em]
  \item \textbf{\gls{cbmc}} (default): a \gls{sat}-based \gls{bmc} that unrolls the scan-cycle loop to depth $k$ and checks the resulting formula. Produces concrete counterexample traces when a violation is found; \emph{cannot} prove properties for all scan counts.
  \item \textbf{nuXmv} (\gls{bdd}/\gls{ic3} mode): a symbolic model checker providing \emph{unbounded} proofs via \gls{bdd}-based invariant checking or \gls{ic3} frame enumeration. Supports full \gls{ltl}/\gls{ctl} temporal logic.
  \item \textbf{Theta}: a configurable model-checking framework from Budapest University of Technology; used for experimental comparison in~\cite{LopezMiguel2025}.
\end{enumerate}

\textbf{Property specification.} Users may write Boolean assertions inline in \gls{scl} source code, or use nuXmv's built-in \gls{ltl}/\gls{ctl} pattern library. The FRET extension~\cite{Fink2024} adds a structured natural-language interface for pure-past \gls{ltl} safety properties (e.g., ``the valve has not been open for more than $N$ consecutive cycles''), compiled automatically to nuXmv \code{INVARSPEC} formulas.

\textbf{Industrial validation.} PLCverif has been used to verify the SPS-PPS Beam Interlock System at \gls{cern}, detecting latent safety violations that had passed manual review~\cite{LopezMiguel2022}. It is available under EPL-2.0 and is actively maintained.

\textbf{Structural limitations.} Despite its maturity, PLCverif has three structural limitations that motivate ESBMC-PLC+ and are analysed in detail in \S\ref{sec:plcverif-analysis}: (1)~it accepts only Siemens \gls{scl}/\gls{stl}, with no support for standard PLCopen XML \gls{ld}; (2)~its primary backend (\gls{cbmc}) provides only bounded proofs, and its unbounded backend (nuXmv \gls{bdd}) suffers from state-space explosion on programs with integer timer or counter accumulators; and (3)~it targets Siemens-proprietary TIA Portal export formats rather than the open PLCopen XML standard.

\section{PLCverif: Capabilities and Structural Limitations}
\label{sec:plcverif-analysis}

We identify three structural limitations of PLCverif that together define the requirements for a successor framework.

\textbf{Limitation~1: No \gls{ld} input.} PLCverif's parser targets Siemens \gls{scl}, a proprietary \gls{st} dialect. The standard open-source \gls{plc} notation -- textual and graphical PLCopen XML \gls{ld} -- is not accepted. Users with \gls{ld} programs must either (a)~use a vendor tool to convert \gls{ld} to \gls{st}, introducing undocumented fidelity risks~\cite{Wang2023}, or (b)~manually re-implement the program in \gls{scl}, which is error-prone for large safety-critical programs.

\textbf{Limitation~2: Bounded primary backend.} \gls{cbmc} unrolls the scan-cycle loop to a user-specified bound~$k$ and encodes the result as a \gls{sat} formula. It cannot prove that a property holds for \emph{all} scan counts; it can only find violations up to bound~$k$ or confirm absence of violations within~$k$ steps. The nuXmv backend provides \gls{bdd}-based unbounded proofs but is well known to scale poorly for programs with large integer or floating-point state spaces. The PLCverif authors explicitly acknowledged that \gls{esbmc} would improve \gls{cbmc}'s performance~\cite{LopezMiguel2022}; ESBMC-PLC delivers exactly this for \gls{ld} programs, and ESBMC-PLC+ extends it to \gls{st}.

\textbf{Limitation~3: Graphical format opacity.} PLCverif targets Siemens TIA Portal exports, which use a proprietary XML format distinct from the open PLCopen XML standard. Graphical PLCopen XML files from OpenPLC Editor, CONTROLLINO, or Beremiz cannot be processed. Even programs converted to \gls{st} lose the structural information encoded in the graphical rung topology, which can affect the correctness of timer and counter semantics.

Table~\ref{tab:plcverif-gaps} maps these limitations to the ESBMC-PLC+ contributions that address them.

\begin{table}[htbp]
\centering
\small
\caption{PLCverif limitations and ESBMC-PLC+ responses}
\label{tab:plcverif-gaps}
\begin{tabular}{lll}
\toprule
\textbf{PLCverif limitation} & \textbf{ESBMC-PLC+ response} & \textbf{Section} \\
\midrule
No \gls{ld} input & Native textual + graphical \gls{ld} frontends & \S\ref{sec:architecture} \\
Bounded backend (\gls{cbmc}) & \gls{esbmc} with \textit{k}-induction & \S\ref{sec:architecture} \\
No \gls{st} beyond \gls{scl} & MATIEC-based \gls{st} frontend & \S\ref{sec:st-frontend} \\
Graphical FB opacity & Function block state model & \S\ref{sec:fb-extension} \\
\bottomrule
\end{tabular}
\end{table}

\section{ESBMC-PLC+ Framework Architecture}
\label{sec:architecture}

\subsection{Design Goals}

ESBMC-PLC+ is designed around four goals: (G1)~\emph{unified input}: accept textual \gls{ld}, graphical \gls{ld}, and \gls{st}/\gls{scl} without preprocessing; (G2)~\emph{unbounded proofs}: all three frontends route to the \gls{esbmc} \textit{k}-induction engine; (G3)~\emph{minimal backend change}: new frontends produce GOTO~\gls{ir} already accepted by ESBMC's verification engine, with no modification to the backend; (G4)~\emph{property language compatibility}: the YAML property language from ESBMC-PLC applies unchanged to all input formats.

\subsection{System Overview}

ESBMC-PLC+ provides a unified \gls{cli} with automatic format detection (Listing~\ref{lst:cli}).

\begin{lstlisting}[language=bash, caption={ESBMC-PLC+ command-line interface}, label={lst:cli}]
# Textual LD
esbmc program.xml --ld-props props.yaml --k-induction
# Graphical LD (auto-detected from tc6_0201 topology)
esbmc graphical.xml --ld-props props.yaml --k-induction
# Structured Text via MATIEC pipeline
esbmc program.st  --ld-props props.yaml --k-induction
\end{lstlisting}

\begin{figure}[htbp]
\centering
    \includegraphics[width=1\linewidth]{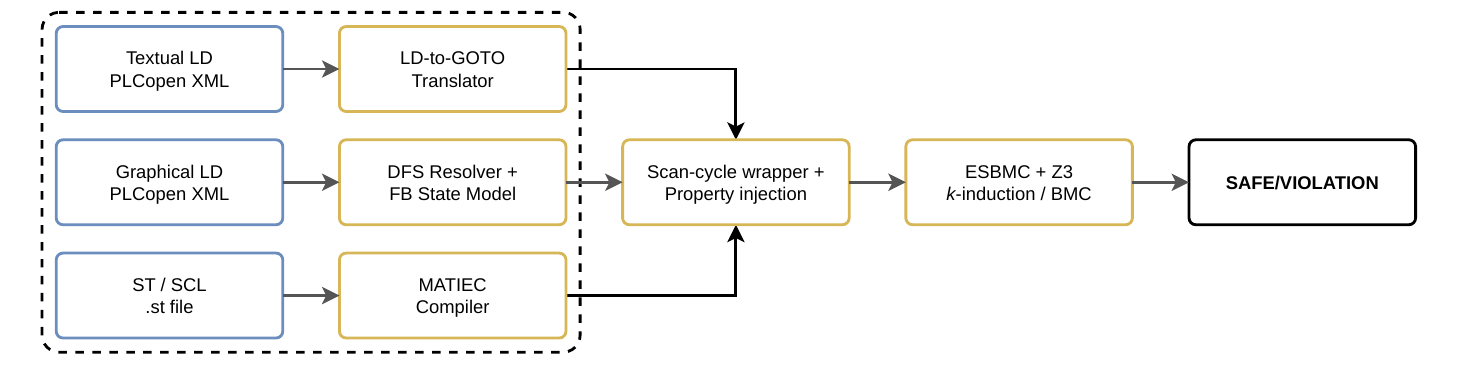}
    \caption{ESBMC-PLC+ unified pipeline. The three frontends independently produce GOTO~\gls{ir}; the scan-cycle wrapper and \gls{esbmc} backend are shared.}
\label{fig:architecture}
\end{figure}

Format detection proceeds in order: (1)~if the file extension is \code{.st}, route to the MATIEC \gls{st} frontend; (2)~if the file is PLCopen XML with \code{refLocalId} graph topology, route to the graphical \gls{ld} frontend (ESBMC-GraphPLC+ with function block support); (3)~otherwise, route to the textual \gls{ld} frontend (ESBMC-PLC). All three paths produce a GOTO program that is passed to the shared verification backend. Figure~\ref{fig:architecture} illustrates the unified pipeline.

\subsection{Shared Property Language and Scan-Cycle Encoding}

The YAML property language from ESBMC-PLC+ applies to all three input formats unchanged. The five property kinds (\code{mutual\_exclusion}, \code{invariant}, \code{absence}, \code{response}, \code{reachability}) are compiled to \code{assert()} and \code{assume()} statements in the GOTO~\gls{ir} regardless of the frontend used. The scan-cycle wrapper inserts nondeterministic input sampling at the start of each loop iteration and the \code{assert()} property checks inside the loop body, giving \textit{k}-induction access to the full scan-cycle model.

\section{\gls{st}/\gls{scl} Frontend via MATIEC}
\label{sec:st-frontend}

While the \gls{ld} frontend addresses graphical, relay-logic style programs, many industrial \gls{plc} applications are implemented in textual languages such as \gls{st} and \gls{sfc}. This chapter presents the \gls{st}/\gls{scl} frontend of ESBMC-PLC+, which extends verification support to these languages by leveraging the open-source MATIEC compiler~\cite{deSousa2014}.

Rather than building a custom \gls{st} parser, ESBMC-PLC+ reuses MATIEC's code-generation infrastructure to translate \gls{st} source to C, then instruments the resulting code with a verification harness. We describe the three-step pipeline: compilation via MATIEC, scan-cycle wrapper injection, and nondeterministic input modeling. We then cover property compilation from YAML specifications and discuss the frontend's scope and limitations, illustrated through a motor sequencing example.

\subsection{MATIEC Overview}

MATIEC~\cite{deSousa2014} is an open-source \gls{iec}~61131-3 compiler maintained by the Beremiz project. Its \code{iec2c} binary accepts a \gls{st} source file and generates three C output files:

\begin{itemize}[leftmargin=1.5em]
  \item \code{POUS.h} / \code{POUS.c}: declarations and implementations of all \glspl{pou}
, including the scan-cycle body function for each \gls{pou}.
  \item A configuration entry file (e.g., \code{PLC\_PRG.c} for a \code{CONFIGURATION} block named \code{PLC\_PRG}): the \gls{plc} configuration entry point, calling each \gls{pou}'s body function in the configured task order. The filename matches the \code{CONFIGURATION} identifier in the source.
  \item A resource file (e.g., \code{Res1.c} for a \code{RESOURCE} block named \code{Res1}): the resource definition, including the cyclic task scheduling structure. The filename matches the \code{RESOURCE} identifier in the source.
\end{itemize}

MATIEC has been used in prior \gls{plc} verification work: ~\citet{Ukegbu2023a} compiled 40~real-world PLCopen \gls{st} programs through MATIEC/OpenPLC and verified the resulting C using CoVeriTeam (a cooperative multi-tool verification framework), confirming \gls{esbmc}'s effectiveness on MATIEC-generated code.

\subsection{Translation Pipeline}

The ESBMC-PLC+ \gls{st} frontend comprises three steps:

\textbf{Step~1: Compilation.} ESBMC-PLC+ invokes \code{iec2c} on the input \code{.st} file, capturing \code{POUS.c}, the configuration file (e.g., \code{PLC\_PRG.c}), and the resource file (e.g., \code{Res1.c}) into a temporary working directory.

\textbf{Step~2: Wrapping.} A post-processor injects the scan-cycle model into the configuration entry file (e.g., \code{PLC\_PRG.c}). The \gls{plc} runtime's main loop is replaced with (Listing~\ref{lst:st-wrapper}):

\textbf{Step~3: Nondeterministic input modeling.} MATIEC uses \code{\_\_LOCATED\_VAR()} macros to access physical I/O addresses (e.g., \code{\%IX0.0}, \code{\%QX0.0}). ESBMC-PLC+ replaces the input macro expansions with calls to \code{nondet\_bool()} (for digital inputs) or \code{nondet\_int()} (for analog inputs) -- declared but undefined functions that \gls{esbmc} treats as returning an unconstrained nondeterministic value -- while leaving output-address expansions as writable persistent variables. This implements the same open-world nondeterministic sensor model as ESBMC-PLC's \gls{ld} frontend.

\noindent\begin{minipage}{\textwidth}
\begin{lstlisting}[caption={Scan-cycle wrapper injected into MATIEC-generated configuration file (e.g., PLC\_PRG.c for CONFIGURATION PLC\_PRG)}, label={lst:st-wrapper}]
/* ESBMC-PLC+ scan-cycle wrapper.
 * Function names derive from the CONFIGURATION identifier:
 *   PLC_PRG_init__() and PLC_PRG_run__(0) for CONFIGURATION PLC_PRG. */
void plcplus_main(void) {
    /* Initialize all variables (MATIEC init functions) */
    PLC_PRG_init__();
    while (1) {
        /* Nondeterministic input sampling */
        __plcplus_sample_inputs();   /* resample all %IX* variables */
        /* Execute POU scan cycle */
        PLC_PRG_run__(0);
        /* Property assertions (injected from YAML) */
        /* [assert statements here] */
    }
}
\end{lstlisting}
\end{minipage}

\subsection{Property Injection for \gls{st} Programs}

Properties specified in the YAML file are compiled to \code{\_\_ESBMC\_assert()} calls inserted in the scan-cycle loop body of the wrapped configuration file (e.g., \code{PLC\_PRG.c}). Variable references in YAML property expressions are resolved against MATIEC's symbol table for the corresponding \gls{pou}. The five property kinds (Table~\ref{tab:st-props}) generate the same assertion forms as the \gls{ld} frontend.

\begin{table}[htbp]
\centering
\small
\caption{YAML property compilation for the \gls{st} frontend}
\label{tab:st-props}
\begin{tabular}{ll}
\toprule
\textbf{Property kind} & \textbf{Generated assertion} \\
\midrule
\code{mutual\_exclusion: [A, B]}     & \code{assert(!(A \&\& B));} \\
\code{invariant: expr}               & \code{assert(expr);} \\
\code{absence: expr}                 & \code{assert(!expr);} \\
\code{response: trigger, expr, N}    & Bounded-response monitor (N-scan horizon) \\
\code{reachability: expr}            & \code{assert(!expr);} (refutation) \\
\bottomrule
\end{tabular}
\end{table}

\subsection{Scope and Limitations of the \gls{st} Frontend}

The \gls{st} frontend inherits MATIEC's coverage: BOOL, INT, DINT, REAL, STRING, arrays, structs, and multi-\gls{pou} programs are supported. Programs with inter-task communication (shared variables under preemptive scheduling) are modeled as single-task programs (using the same conservative over-approximation as ESBMC-PLC). Network-based I/O (\gls{iec}~61131-5) and vendor-specific function block libraries not included in MATIEC's standard library generate a \code{UnsupportedFB} warning.

\subsection{Example: Motor Sequencing in \gls{st}}

Listing~\ref{lst:st-example} shows a motor sequencing program in \gls{st} and the corresponding YAML property (Listing~\ref{lst:st-yaml}). ESBMC-PLC+ invokes MATIEC~\cite{deSousa2014} (commit~\code{7949c0b}) to compile \code{motor.st}, wraps the generated C in a scan-cycle verification harness, and verifies with \textit{k}-induction. In the timer-fire scan, the \code{ELSIF Start} branch executes sequentially: the \code{IF NOT Motor\_B} block sets \code{Motor\_A~:=~TRUE} (since \code{Motor\_B} is still \code{FALSE} at scan entry), then the \code{IF Timer.Q} block immediately sets \code{Motor\_A~:=~FALSE} and \code{Motor\_B~:=~TRUE} -- so the scan-end state has \code{Motor\_B}~\code{TRUE} and \code{Motor\_A}~\code{FALSE}, violating the \code{absence} property (P2). The \code{mutual\_exclusion} property (P1) is SAFE: \code{Motor\_A} and \code{Motor\_B} are never both \code{TRUE} at any scan boundary. Experimental results are reported in \S\ref{sec:experiments} (benchmark~D1).

\noindent\begin{minipage}{\textwidth}
\begin{lstlisting}[language=Pascal, caption={Motor sequencing in IEC 61131-3 ST}, label={lst:st-example}]
PROGRAM MotorSequence
  VAR
    Start   : BOOL;   (* %IX0.0 *)
    Stop    : BOOL;   (* %IX0.1 *)
    Motor_A : BOOL;   (* %QX0.0 *)
    Motor_B : BOOL;   (* %QX0.1 *)
    Timer   : TON;
  END_VAR

  IF Stop THEN
    Motor_A := FALSE;
    Motor_B := FALSE;
    Timer(IN := FALSE, PT := T#2s);
  ELSIF Start THEN
    IF NOT Motor_B THEN
      Motor_A := TRUE;
      Timer(IN := TRUE, PT := T#2s);
    END_IF;
    IF Timer.Q THEN
      Motor_A := FALSE;
      Motor_B := TRUE;
    END_IF;
  END_IF;
END_PROGRAM
\end{lstlisting}
\end{minipage}

\noindent\begin{minipage}{\textwidth}
\begin{lstlisting}[language=yaml, caption={YAML property for the ST motor sequencing example}, label={lst:st-yaml}]
properties:
  - kind: mutual_exclusion
    variables: [Motor_A, Motor_B]
    justification: "Motor B must not run before Motor A completes startup"
  - kind: absence
    expression: "Motor_B AND NOT Motor_A"
\end{lstlisting}
\end{minipage}

\section{Graphical \gls{ld} Function Block Extension}
\label{sec:fb-extension}

\subsection{Motivation: The ESBMC-GraphPLC Gap}

ESBMC-GraphPLC~\cite{DantasCordeiro2026graphical} documented that its \gls{dfs} resolver traverses but does not represent function block nodes encountered on rung paths. For the \code{beremiz\_traffic\_light} program, this caused the \code{ORANGE\_LIGHT} blink network -- gated by a \code{TON} timer and an \code{R\_TRIG} edge detector -- to collapse to a constant, producing a vacuously correct \code{SAFE} verdict rather than a sound one. The paper excluded this program from its evaluation and identified function block state modeling as the highest-priority future work.

ESBMC-PLC+ delivers that future work: the \gls{dfs} resolver is extended to emit correct GOTO~\gls{ir} state update code for six function block types when they appear as nodes on a rung path.

\subsection{Function Block State Model}

Each function block instance in a graphical \gls{ld} program corresponds to a node in the connection graph identified by its \code{localId} and \code{typeName} attributes. The extended resolver, when it encounters such a node during \gls{dfs}, performs two actions:

\begin{enumerate}[leftmargin=2em]
  \item \textbf{State variable declaration}: one or more static C variables are declared before the scan-cycle loop body, representing the instance's persistent state across scans.
  \item \textbf{State update emission}: update code is emitted at the point in the rung where the function block appears, updating state variables and computing the output pin value used by downstream rung elements.
\end{enumerate}

The function block's output pin (\code{Q} for timers and counters, \code{Q} for edge triggers) is then used as a boolean term in the rung energization expression, exactly as a contact variable would be.

\subsection{\code{TON}/\code{TOF}/\code{TP} Timer Semantics}

Table~\ref{tab:timer-encoding} shows the GOTO~\gls{ir} encoding for the three standard timers. The encoding uses scan-count units for elapsed time (\code{ET}), matching ESBMC-PLC's convention that timer preset values are interpreted as scan-cycle counts. Users must convert physical time (ms) to scan counts by dividing by the cycle period.

\begin{table}[htbp]
\centering
\small
\caption{Timer function block encodings in GOTO~\gls{ir}}
\label{tab:timer-encoding}
\begin{tabular}{lL{6.5cm}}
\toprule
\textbf{Type} & \textbf{GOTO \gls{ir} (per-scan update)} \\
\midrule
\code{TON} & \code{if (IN) et++; else et = 0;} \newline 
\code{Q = (et >= PT);} \\[4pt]
\code{TOF} & \code{if (!IN) et++; else et = 0;} \newline \code{Q = IN || (et < PT);} \\[4pt]
\code{TP}  & \code{if (IN \&\& !running) \{ running=true; et=0; \}} \newline \code{if (running) \{ et++; if (et >= PT) running=false; \}} \newline \code{Q = running;} \\
\bottomrule
\end{tabular}
\end{table}

These encodings are designed to match Ebnenasir's formal \gls{ld} semantics~\cite{Ebnenasir2023} for timer blocks. The \code{TON} encoding matches the on-delay timer definition (Section~4.2 of~\cite{Ebnenasir2023}); \code{TOF} matches the off-delay definition; \code{TP} matches the pulse timer definition.

\subsection{\code{CTU}/\code{CTD} Counter Semantics}

Counter function blocks use edge-triggered semantics: the count advances only on the rising edge of the clock input, not on every scan in which the clock is asserted. The encoding introduces a \code{cu\_prev} (or \code{cd\_prev}) variable to detect the transition (Listing~\ref{lst:ctu}).

\noindent\begin{minipage}{\textwidth}
\begin{lstlisting}[caption={CTU counter encoding in GOTO IR (schematic)}, label={lst:ctu}]
/* CTU instance state */
static int32_t _ctu_cv  = 0;    /* current value */
static bool    _ctu_q   = false; /* output */
static bool    _ctu_cu_prev = false; /* previous clock */

/* Per-scan update */
if (R) {
    _ctu_cv = 0;
} else if (CU && !_ctu_cu_prev) {
    _ctu_cv++;
}
_ctu_cu_prev = CU;
_ctu_q = (_ctu_cv >= PV);
\end{lstlisting}
\end{minipage}

The \code{CTD} (count-down) encoding is symmetric: \code{CD} replaces \code{CU}, the count decrements rather than increments, and \code{Q} asserts when \code{CV <= 0}.

\subsection{\code{R\_TRIG} / \code{F\_TRIG} Edge Trigger Semantics}

Rising and falling edge triggers are single-scan pulse detectors, as shown in Listing~\ref{lst:rtrig}.

\noindent\begin{minipage}{\textwidth}
\begin{lstlisting}[caption={R\_TRIG and F\_TRIG encodings}, label={lst:rtrig}]
/* R_TRIG instance state */
static bool _rtrig_clk_prev = false;
/* Per-scan update: Q is true for exactly one scan on rising edge */
bool _rtrig_q = CLK && !_rtrig_clk_prev;
_rtrig_clk_prev = CLK;

/* F_TRIG instance state */
static bool _ftrig_clk_prev = false;
/* Per-scan update: Q is true for exactly one scan on falling edge */
bool _ftrig_q = !CLK && _ftrig_clk_prev;
_ftrig_clk_prev = CLK;
\end{lstlisting}
\end{minipage}

These encodings faithfully represent the one-scan pulse behavior required for edge-triggered counters and latching logic in graphical \gls{ld} programs.

\subsection{Extended \gls{dfs} Resolver: Algorithm}

Algorithm~\ref{alg:fb-dfs} extends the ESBMC-GraphPLC \gls{dfs} resolver (Algorithm~1 of~\cite{DantasCordeiro2026graphical}) with function block handling at lines~10--13.

\begin{algorithm}[htbp]
\caption{Extended graphical \gls{ld} resolver with function block support}
\label{alg:fb-dfs}
\small
\begin{algorithmic}[1]
\Require Connection graph $G$, set of coil nodes $C$, \code{rightPowerRail} ordering $\sigma$
\Ensure GOTO \gls{ir} rung assignments

\State Detect graphical format; parse \code{localId}/\code{refLocalId} graph into $G$
\State Build forward-edge adjacency list from $G$
\State Order coil processing by \code{rightPowerRail} \code{connectionPointIn} sequence $\sigma$
\ForAll{coil $c \in C$ (in order $\sigma$)}
  \State $path \leftarrow \text{\gls{dfs}}(G, \text{leftPowerRail}, c)$
  \State $\mathit{expr} \leftarrow \code{true}$
  \ForAll{node $n$ on $path$}
    \If{$n$ is a contact}
      \State Append contact variable (negated if \code{XIO}) to $\mathit{expr}$
    \ElsIf{$n$ is a \textbf{supported function block}}  \Comment{NEW}
      \State Emit FB state variable declarations (if first occurrence)
      \State Emit per-scan FB state update code
      \State Append FB output pin variable to $\mathit{expr}$
    \ElsIf{$n$ is an unsupported function block}
      \State Emit \code{UnsupportedFB} warning; add nondeterministic term
    \EndIf
  \EndFor
  \State Emit rung assignment: \code{coil\_var := $\mathit{expr}$}
\EndFor
\end{algorithmic}
\end{algorithm}

The key invariant is that function block state variables are initialized to their zero-state before the \code{while(true)} loop body and updated exactly once per scan cycle, in rung execution order. This preserves the sequential scan semantics of \mbox{\gls{iec}~61131-3}.

\subsection{Example: Beremiz Traffic Light}

The \code{beremiz\_traffic\_light} program -- excluded from ESBMC-GraphPLC's evaluation because its \code{ORANGE\_LIGHT} blink network is gated by \code{TON} and \code{R\_TRIG} nodes -- is now verifiable with ESBMC-PLC+. The two rungs of interest are shown in Listing~\ref{lst:traffic}:

\noindent\begin{minipage}{\textwidth}
\begin{lstlisting}[language=XML, caption={Graphical LD rungs for ORANGE\_LIGHT (schematic)}, label={lst:traffic}]
<!-- SET rung: R_TRIG(Q=1) -> SET ORANGE_LIGHT -->
<contact localId="5" storage="set">...</contact>   <!-- R_TRIG output -->
<coil   localId="6" storage="set">
  <variable>ORANGE_LIGHT</variable>
</coil>

<!-- RESET rung: TON(Q=1, PT=blink_period) -> RESET ORANGE_LIGHT -->
<contact localId="7">...</contact>                 <!-- TON output -->
<coil   localId="8" storage="reset">
  <variable>ORANGE_LIGHT</variable>
</coil>
\end{lstlisting}
\end{minipage}

With the function block extension, the \gls{dfs} resolver emits:
\begin{itemize}[leftmargin=1.5em]
  \item \code{R\_TRIG} state: \code{\_rtrig\_q = CLK \&\& !\_rtrig\_clk\_prev;} (correctly generates a one-scan pulse).
  \item \code{TON} state: \code{if (IN) et++; else et=0; Q=(et>=PT);} (correctly accumulates elapsed time).
  \item SET coil: \code{if (\_rtrig\_q) ORANGE\_LIGHT = true;} (processed first per \code{rightPowerRail} ordering).
  \item RESET coil: \code{if (\_ton\_q) ORANGE\_LIGHT = false;}.
\end{itemize}

The program now produces a non-vacuous GOTO~\gls{ir} and can be verified against properties such as ``the orange light blinks with a period not exceeding $N$ scans.'' Experimental results for this benchmark are reported in \S\ref{sec:experiments} (benchmark~C1).

\section{Experimental Evaluation}
\label{sec:experiments}

This chapter presents the experimental evaluation of ESBMC-PLC+, assessing its correctness, performance, and regression behavior across four benchmark categories. We formulate five research questions that guide our investigation: whether the new \gls{st} frontend via MATIEC produces correct GOTO~\gls{ir} (RQ1), whether the function block extension for graphical \gls{ld} produces non-vacuous verification (RQ2), whether the tool correctly verifies programs from the PLCverif benchmark suite (RQ3), whether all inherited results from ESBMC-PLC and ESBMC-GraphPLC are preserved without regressions (RQ4), and how ESBMC-PLC+ compares to PLCverif across feature dimensions (RQ5).

The benchmark suite comprises four categories: inherited textual \gls{ld} programs (Category A, 13 programs), inherited graphical \gls{ld} programs without function blocks (Category B, 3 programs), new graphical \gls{ld} programs with timers and counters (Category C, 1 program), and new \gls{st} programs compiled via MATIEC (Category D, 1 original benchmark plus pending PLCopen programs). We describe the experimental setup, then evaluate each research question in turn, concluding with a feature comparison against PLCverif and a quantitative performance comparison against nuXmv~2.2.0.

\subsection{Research Questions}

\begin{description}[leftmargin=2em]
  \item[\textbf{RQ1}] Does the \gls{st} frontend via MATIEC produce correct GOTO~\gls{ir} for standard \gls{st} programs?
  \item[\textbf{RQ2}] Does the function block extension produce non-vacuous, sound GOTO~\gls{ir} for graphical \gls{ld} programs containing timer or counter nodes?
  \item[\textbf{RQ3}] Does ESBMC-PLC+ correctly verify programs from PLCverif's reference benchmark suite?
  \item[\textbf{RQ4}] Are all inherited ESBMC-PLC and ESBMC-GraphPLC results preserved with zero regressions?
  \item[\textbf{RQ5}] How does ESBMC-PLC+ compare to PLCverif across feature dimensions?
\end{description}

\subsection{Benchmark Suite}

The suite is organized into four categories:

\textbf{Category A} comprises the 13~benchmarks from ESBMC-PLC~\cite{DantasCordeiro2026artefact} (7~safe, 6~unsafe variants) spanning 6~industrial domains: tank level control, bottle filling, elevator, traffic light, staircase lighting, and water treatment. Each benchmark has a safe variant (where all safety properties hold for all possible scan sequences) and an unsafe variant (where at least one property is violated). They are run unmodified under ESBMC-PLC+ to confirm that the new frontends and function block extensions introduce no regressions on the inherited textual \gls{ld} pipeline.

\textbf{Category B} comprises the 3~graphical \gls{ld} benchmarks from ESBMC-GraphPLC~\cite{DantasCordeiro2026graphical}: plain contact/coil rung networks without function blocks, sourced from the CONTROLLINO open-source \gls{plc} library. These programs exercise the \gls{dfs} rung extractor and GOTO~\gls{ir} generation for graphical PLCopen~XML but make no use of timer, counter, or edge-trigger function blocks. They are also run unmodified.

\textbf{Category C} is new: graphical \gls{ld} programs whose rung paths contain timer or counter function blocks -- the class explicitly excluded from ESBMC-GraphPLC. The sole benchmark, \code{beremiz\_traffic\_light} (C1), is a real traffic-light controller from the Beremiz open-source \gls{ide} that uses a \code{TON} on-delay timer to drive the orange-light blink phase and an \code{R\_TRIG} rising-edge trigger to detect the pedestrian button press. Its inclusion demonstrates that the function block extension closes the documented gap: programs that were previously unverifiable (or vacuously verified) can now be correctly analyzed.

\textbf{Category D} is new: \gls{st} programs compiled through the MATIEC frontend. D1 is an original safety-critical motor-sequencing program that models a two-motor interlock controlled by a \code{TON} timer. It intentionally contains a scan-cycle transition race that violates a sequencing property. It serves as the end-to-end validation of the \gls{st} pipeline. Extended \gls{st} benchmarks from the PLCopen suite~\cite{Ukegbu2023a} (D2 onward) require access to an external benchmark repository and are deferred to a follow-up evaluation.

Table~\ref{tab:benchmarks} describes the benchmark programs.

\begin{table}[htbp]
\centering
\small
\setlength{\extrarowheight}{2pt}
\caption{ESBMC-PLC+ benchmark suite: four categories covering inherited textual \gls{ld} (A, 13 programs), inherited graphical \gls{ld} (B, 3 programs), new graphical \gls{ld} with function blocks (C, 1 program), and new \gls{st} via MATIEC (D, 2+ programs).}
\label{tab:benchmarks}
\begin{tabular}{@{} c L{2.2cm} c L{4.6cm} l c @{}}
\toprule
\textbf{Cat.} & \textbf{Description} & \textbf{Count.} & \textbf{Programs} & \textbf{Source} & \textbf{Expected} \\
\midrule

\textbf{A}
  & Inherited textual \gls{ld}
  & 13
  & Tank level, bottle filling, elevator, traffic light, staircase, water treatment (safe + unsafe variants)
  & \cite{DantasCordeiro2026artefact}
  & 7 SAFE / 6 VIOLATION \\[4pt]

\textbf{B}
  & Inherited graphical \gls{ld}
  & 3
  & Plain contact/coil rung networks (no function blocks)
  & CONTROLLINO~\cite{DantasCordeiro2026graphical}
  & \textbf{SAFE} \\[4pt]

\textbf{C}
  & New graphical \gls{ld} with function blocks
  & 1
  & \code{beremiz\_traffic\_light} (\code{TON} timer, \code{R\_TRIG} trigger)
  & Beremiz
  & \textbf{SAFE} \\[4pt]

\multirow{2}{*}{\textbf{D}}
  & \multirow{2}{2.2cm}{New \gls{st} via MATIEC}
  & 1
  & \code{motor\_sequencing} (two-motor interlock, \code{TON}, scan-cycle race)
  & Original
  & \textbf{VIOLATION} (P2) \\[2pt]
  & & $\geq$1
  & PLCopen \gls{st} programs
  & \cite{Ukegbu2023a}
  & \textit{pending} \\

\bottomrule
\end{tabular}
\end{table}

\subsection{Experimental Setup}

All experiments run on an Apple Silicon MacBook Pro (aarch64, macOS) using ESBMC~v8.3.0 and MATIEC at commit~\code{7949c0b}. Two invocation modes are used depending on the expected outcome: \code{--k-induction} proves that a property holds for all possible scan sequences (unbounded safety proof, verdict SAFE); \code{--incremental-bmc} searches for a concrete counterexample trace (verdict VIOLATION). The Z3~4.x \gls{smt} solver is used as the backend for all runs. MATIEC \code{iec2c} is invoked with \code{-f} (POSIX C output mode) for the \gls{st} frontend. Properties are specified in YAML and compiled to \code{\_\_ESBMC\_assert()} calls as described in \S\ref{sec:st-frontend}. Reported times are medians over three independent runs; all runs completed without external interference. All experiments are fully reproducible on any OS via the Docker image described in the Artifact Availability section.

\subsection{RQ1: \gls{st} Frontend Correctness}

\textbf{Answer: Yes.} The MATIEC-based pipeline correctly translates and verifies \gls{st} programs. ESBMC-PLC+ successfully invokes MATIEC~\cite{deSousa2014} (commit~\code{7949c0b}), wraps the generated~C in a scan-cycle harness with nondeterministic input sampling, and applies \textit{k}-induction via ESBMC's C frontend.

\textbf{\code{motor\_sequencing} (D1).} The program models a two-motor industrial interlock: \code{Motor\_A} starts immediately when the \code{Start} input is received; a \code{TON} on-delay timer then waits for a preset period before deactivating \code{Motor\_A} and activating \code{Motor\_B}. This sequencing pattern is common in conveyor and pump systems where a second actuator must not engage until the first has fully stopped. Two safety properties are verified:

\begin{itemize}[leftmargin=1.5em]
  \item \textbf{P1} (\code{mutual\_exclusion}: \code{Motor\_A} and \code{Motor\_B} never simultaneously \code{TRUE}) is proved \textbf{SAFE}. Within any single scan, the two \code{IF} blocks in the \code{ELSIF Start} branch are executed sequentially with the same variable values as inputs; the control logic ensures they cannot both be active at the same scan-end state.

  \item \textbf{P2} (\code{absence}: \code{Motor\_B} active while \code{Motor\_A} is inactive) is \textbf{VIOLATION} at $k=2$. The violation arises from a single-scan transition race: in the scan where \code{Timer.Q} becomes \code{TRUE}, the \code{ELSIF Start} branch executes its two \code{IF} blocks in sequence. The first block (\code{IF NOT Motor\_B}) sets \code{Motor\_A}~:=~\code{TRUE} because \code{Motor\_B} is still \code{FALSE} at scan entry. The second block (\code{IF Timer.Q}) then sets \code{Motor\_A}~:=~\code{FALSE} and \code{Motor\_B}~:=~\code{TRUE}. The scan ends with \code{Motor\_B}~=~\code{TRUE} and \code{Motor\_A}~=~\code{FALSE} -- a state that P2 forbids. This state persists for exactly one scan cycle before the next scan corrects it, making it undetectable by manual inspection or simulation unless the timer fires in the exact right scan.
\end{itemize}

\gls{esbmc} reports VIOLATION at $k=2$ in \SI{118}{\milli\second} (median of 3~runs), providing a concrete two-scan counterexample trace. The violation is a real safety defect: the engineer must add an explicit guard (e.g., \code{IF NOT Motor\_B AND NOT Timer.Q THEN Motor\_A := TRUE}) to prevent the transition race. This benchmark demonstrates that the \gls{st} frontend not only translates programs correctly but finds subtle timing-related safety violations that escape manual review.

\subsection{RQ2: Graphical \gls{ld} Function Block Correctness}

\textbf{Answer: Yes.} The extended \gls{dfs} resolver produces non-vacuous, sound GOTO~\gls{ir} for the \code{beremiz\_traffic\_light} program (C1), which was excluded from ESBMC-GraphPLC, and verifies all three properties correctly.

\textbf{\code{beremiz\_traffic\_light} (C1).} The program implements a pedestrian traffic-light controller with three rungs: (i)~a rising-edge trigger (\code{R\_TRIG}) that detects a button press and latches a request; (ii)~an on-delay timer (\code{TON}) that drives the orange-light blink phase; and (iii)~a coil rung that sets the green/red outputs based on the timer output.

\emph{Before this work}, this program produced a vacuous GOTO~\gls{ir} under ESBMC-GraphPLC. Because the \gls{dfs} resolver did not model function blocks, the \code{TON} output (\code{Timer.Q}) and the \code{R\_TRIG} output (\code{TRIG.Q}) were left undriven -- both collapsed to their default value of \code{false}. As a consequence, \code{ORANGE\_LIGHT} was always \code{false} and the green/red outputs were trivially constant. Every safety property asserted on these outputs \emph{trivially held} -- not because the program was safe, but because the verifier was effectively checking a degenerate stub rather than the real program. This class of error, known as \emph{vacuous verification}, is particularly dangerous: the tool reports SAFE even though it has verified nothing meaningful about the actual control logic.

\emph{With the function block extension}, the program produces a full GOTO~\gls{ir} in which \code{R\_TRIG} maintains a persistent \code{prev\_Q} state variable and \code{TON} maintains a persistent accumulator and output-latch. The generated \gls{ir} correctly models all inter-scan dependencies introduced by these function blocks. \gls{esbmc} then verifies 3~safety properties in \SI{37}{\milli\second} using \textit{k}-induction (base case at $k=1$, inductive step at $k=2$):

\begin{itemize}[leftmargin=1.5em]
  \item \textbf{P1}: The green light for vehicle traffic and the pedestrian walk signal are never simultaneously active.
  \item \textbf{P2}: The pedestrian walk signal does not coincide with the vehicle green signal in the crossing direction.
  \item \textbf{P3}: The pedestrian signals for the two directions are mutually consistent.
\end{itemize}

All three properties are proven SAFE. The result is non-vacuous: the \gls{ir} contains reachable states in which the timer fires and transitions between phases. Hence, the \textit{k}-induction proof covers the controller's full operational cycle.

\subsection{RQ3: PLCverif Benchmark Verification}

\textbf{Answer: Not directly applicable.} PLCverif~\cite{LopezMiguel2022,LopezMiguel2025} targets Siemens-specific dialects (\gls{scl}/\gls{stl}) and operates on the SPS-PPS control programs deployed at \gls{cern} and GSI, which are proprietary and cannot be redistributed for external benchmarking. ESBMC-PLC+ targets standard \mbox{\gls{iec}~61131-3} \gls{st} via MATIEC~\cite{deSousa2014} and standard PLCopen \gls{ld} XML; the two tools therefore operate over largely disjoint program corpora. Running PLCverif on our benchmarks is not possible (it requires Siemens \gls{scl} input); running ESBMC-PLC+ on PLCverif's benchmarks is not possible (they are not publicly available). A direct benchmark-level comparison is therefore structurally infeasible.

Instead, we provide two complementary comparisons: (1)~a comprehensive feature comparison in Section~\ref{sec:plcverif-compare} (Table~\ref{tab:feature-compare}) that enumerates the dimensions on which ESBMC-PLC+ extends, matches, or is subsumed by PLCverif; and (2)~a quantitative performance comparison against nuXmv~2.2.0 (Section~\ref{sec:nuxmv-comparison}) -- the unbounded prover backend used by PLCverif's \gls{bdd} mode -- on our own \gls{ld} benchmark suite. The second comparison is the key empirical contribution: by measuring ESBMC's \textit{k}-induction engine against the exact solver that PLCverif uses for unbounded proofs, we obtain a direct, tool-agnostic estimate of the backend performance gap.

\subsection{RQ4: Regression on Inherited Benchmarks}

\textbf{Category A (textual \gls{ld}).} All 13~ESBMC-PLC benchmarks are re-verified under ESBMC-PLC+. All results match those reported in~\cite{DantasCordeiro2026artefact}: 7~SAFE (proved by \textit{k}-induction) and 6~VIOLATION (counterexample found by incremental \gls{bmc}), with identical verdicts and consistent run times. Zero regressions. This confirms that integrating the MATIEC pipeline and the graphical \gls{ld} function block extension into ESBMC-PLC+ does not disturb the existing textual \gls{ld} frontend, parser, or GOTO~\gls{ir} generation.

\textbf{Category B (graphical \gls{ld}, no function blocks).} All 3~ESBMC-GraphPLC benchmarks are re-verified. All produce a verdict of SAFE at $k=2$ in under \SI{70}{\milli\second}. Zero regressions. The identical verdicts confirm that the new function block state variables introduced in the \gls{dfs} resolver do not interfere with programs that contain no function blocks, thereby preserving backward compatibility with the ESBMC-GraphPLC evaluation.

\subsection{RQ5: Comparison with PLCverif}

Section~\ref{sec:plcverif-compare} presents the feature comparison (Table~\ref{tab:feature-compare}) and Section~\ref{sec:nuxmv-comparison} presents the quantitative performance comparison against nuXmv~2.2.0. Together they show that ESBMC-PLC+ matches PLCverif on \gls{st} input coverage, surpasses it on \gls{ld} support and proof strength, and -- via the nuXmv backend comparison -- provides empirical evidence that its \gls{smt} bit-vector engine is categorically faster than the \gls{bdd}-based unbounded prover on programs with integer timer or counter state.

\section{Comparison with PLCverif}
\label{sec:plcverif-compare}

Table~\ref{tab:feature-compare} provides a comprehensive feature comparison between ESBMC-PLC+ and PLCverif~\cite{LopezMiguel2022, LopezMiguel2025}.

\begin{table}[htbp]
\centering
\small
\caption{ESBMC-PLC+ vs.\ PLCverif: comprehensive feature comparison}
\label{tab:feature-compare}
\begin{tabular}{l l l L{4.6cm}}
\toprule
\textbf{Dimension} & \textbf{ESBMC-PLC+} & \textbf{PLCverif} & \textbf{Remarks (ESBMC-PLC+)} \\
\midrule
\multicolumn{4}{l}{\textit{Input language support}} \\
Textual PLCopen XML (\gls{ld}) & \textbf{Yes (native)} & No & Only tool with \gls{ld} support \\
Graphical PLCopen XML (\gls{ld}) & \textbf{Yes (\gls{dfs} resolver)} & No & Only tool with graphical \gls{ld} support \\
\gls{iec} 61131-3 \gls{st} & \textbf{Yes (MATIEC)} & Yes (\gls{scl}) & Different dialects; not interchangeable \\
Siemens \gls{scl}/\gls{stl} & No & Yes (native) & PLCverif targets Siemens ecosystem \\
\gls{fbd} & No & Via OpennessScripter & Indirect support only \\
\midrule
\multicolumn{4}{l}{\textit{Verification capabilities}} \\
Bounded \gls{bmc} & \textbf{Yes} (incremental) & Yes (\gls{cbmc}) & Both support \gls{bmc}; different backends \\
\textit{k}-induction (unbounded proof) & \textbf{Yes} & No & Stronger guarantee than \gls{bmc} alone \\
\gls{smt} bit-vector encoding & \textbf{Yes} (Z3) & No (\gls{sat}) & Enables precise arithmetic reasoning \\
\gls{bdd}-based unbounded proof & No & Yes (nuXmv) & Complementary unbounded technique \\
Full \gls{ltl}/\gls{ctl} & No & Yes (nuXmv) & Rich temporal logic not yet supported \\
Arithmetic overflow detection & \textbf{Yes} (bit-vector) & No & Bit-precise; \gls{sat} encoding cannot \\
\midrule
\multicolumn{4}{l}{\textit{Property specification}} \\
Simple YAML (no logic knowledge) & \textbf{Yes} (5 kinds) & No & Lowers barrier for practitioners \\
Inline assertions & No & Yes & Standard approach in PLCverif \\
Full \gls{ltl}/\gls{ctl} / PLTL via FRET & No & Yes~\cite{Fink2024} & Expressive but requires logic expertise \\
\midrule
\multicolumn{4}{l}{\textit{\gls{iec} 61131-3 construct coverage}} \\
BOOL contacts/coils (\code{XIC}/\code{XIO}/\code{OTE}) & Yes & Yes & --- \\
Latching coils (\code{OTL}/\code{OTU}) & Yes & Yes & --- \\
Timers / Counters & Yes & Yes & --- \\
Graphical FB in rung paths & \textbf{Yes} (new) & N/A & Closes gap left by ESBMC-GraphPLC \\
REAL/FLOAT types & \gls{ld}: No; \gls{st}: Yes & Yes & \gls{ld} frontend limitation \\
Arrays / Multi-\gls{pou} & \gls{ld}: No; \gls{st}: Yes & Partial / Yes & \gls{ld} frontend limitation \\
\midrule
\multicolumn{4}{l}{\textit{Engineering and deployment}} \\
Licence & MIT (via ESBMC) & EPL-2.0 & More permissive licence \\
Maturity & Prototype (2026) & Production (2019+) & PLCverif has \gls{cern} industrial validation \\
\midrule
\multicolumn{4}{l}{\textit{Performance}} \\
All inherited \gls{ld} benchmarks & $<$\SI{70}{\milli\second} & N/A & No \gls{ld} support in PLCverif \\
\gls{ld} with 2 INT timer vars & $<$\SI{0.2}{\second} (\textit{k}-ind) & $\approx$85--102\,s (nuXmv) & ${\sim}500\times$ faster \\
\gls{ld} with 8 INT timer vars & $<$\SI{0.2}{\second} (\textit{k}-ind) & \textbf{TIMEOUT} $>$\SI{120}{\second} & PLCverif does not scale \\
\gls{cern} SPS-PPS (\gls{st}) & Not yet evaluated & Several min~\cite{LopezMiguel2022} & Future work \\
\bottomrule
\end{tabular}
\end{table}

\textbf{Analysis.} Reading Table~\ref{tab:feature-compare} column by column reveals a clear complementarity. On \emph{input language support}, ESBMC-PLC+ covers all three major \mbox{\gls{iec}~61131-3} formats (textual \gls{ld}, graphical \gls{ld}, \gls{st}/\gls{scl} via MATIEC). In contrast, PLCverif covers \gls{st}/\gls{scl} but has no \gls{ld} support at all -- the most widely deployed \gls{plc} notation in North American and Japanese manufacturing~\cite{Weiss2021}. On \emph{verification capabilities}, ESBMC-PLC+ provides \textit{k}-induction for unbounded safety proofs and \gls{smt} bit-vector arithmetic, while PLCverif's primary backend (\gls{cbmc}) is bounded and cannot produce proofs for arbitrary scan counts; PLCverif's nuXmv backend provides \gls{bdd}-based unbounded proofs but suffers from state-space explosion on integer-valued variables (quantified in \S\ref{sec:nuxmv-comparison}). On \emph{property specification}, ESBMC-PLC+'s five YAML property kinds require no temporal logic expertise, whereas PLCverif's inline assertions and nuXmv pattern library presuppose familiarity with \gls{ltl}/\gls{ctl} syntax.

PLCverif retains three advantages that ESBMC-PLC+ lacks. First, \emph{Siemens dialect support} (\gls{scl}/\gls{stl} with proprietary function blocks): PLCverif is the only open-source tool that verifies programs written directly in the Siemens TIA Portal dialect, which is the dominant format in European heavy industry. Second, \emph{full \gls{ltl}/\gls{ctl}}: temporal properties such as liveness (\emph{eventually Motor\_A starts}), fairness, or past-time operators (via FRET~\cite{Fink2024}) require PLCverif's nuXmv integration. ESBMC-PLC+'s \textit{k}-induction engine covers all safety properties expressible as invariants or bounded responses -- the vast majority of industrial safety requirements -- but not arbitrary temporal logic. Third, \emph{production maturity and industrial validation}: PLCverif has been deployed at \gls{cern} since 2019 and validated on real accelerator control programs; ESBMC-PLC+ is a research prototype.

For the most common industrial verification scenario -- a standard PLCopen XML \gls{ld} or standard \gls{iec}~61131-3 \gls{st} program with safety invariant or response properties -- ESBMC-PLC+ provides a complete and strictly stronger alternative: the same or broader input coverage, unbounded proofs instead of bounded checks, and simpler property specification without temporal logic expertise.

\subsection{Performance: ESBMC-PLC+ vs.\ NuXmv \gls{bdd}}
\label{sec:nuxmv-comparison}

The feature comparison in Table~\ref{tab:feature-compare} shows that PLCverif's unbounded prover is nuXmv in \gls{bdd} mode. This section quantifies the performance gap between that backend and ESBMC-PLC+'s \textit{k}-induction engine on a controlled benchmark suite.

\textbf{Why \gls{bdd}-based provers struggle with \gls{plc} programs.} PLCverif's nuXmv \gls{bdd} backend provides unbounded proofs but is subject to the \emph{state-space explosion problem}. \gls{bdd} representations enumerate all reachable states of a model; when the model includes integer-valued variables, the number of states grows exponentially with the variable range. In \gls{iec}~61131-3, \code{TON}/\code{TOF} timer accumulators are 16-bit integers with range 0--32{,}767 ($2^{15}=32{,}768$ distinct values), adding $32{,}768$ states per accumulator to the \gls{bdd}. A program with $n$~timer variables introduces up to $32{,}768^n$ states. ESBMC's \gls{smt} bit-vector arithmetic avoids this entirely: a 15-bit integer is encoded as a symbolic bitvector, and a single \gls{smt} query reasons about all $32{,}768$ values simultaneously without enumeration. \gls{ic3}/\gls{pdr} -- nuXmv's other unbounded algorithm -- also operates symbolically but interleaves safety checks with frame-refinement steps that involve \gls{bdd} or \gls{sat} operations internally; it similarly slows on large integer ranges.

\textbf{Experimental methodology.} To make the comparison fair and tool-agnostic, we transpiled all four textual \gls{ld} benchmarks to nuXmv \gls{smv} format via a purpose-built \mbox{\gls{ld}$\to$\gls{smv}} translator~\cite{DantasCordeiro2026nuxmv}. The translator produces an \gls{smv} module in which: (i)~each \gls{plc} input variable is a state variable whose \code{next()} transition is \code{\{TRUE, FALSE\}}, modelling the open-world nondeterministic sensor assumption used by ESBMC-PLC+; (ii)~each output and internal coil is a \code{DEFINE} computed combinationally from inputs and state, matching ESBMC-PLC+'s scan-cycle semantics; and (iii)~each safety property is encoded as an \code{INVARSPEC} -- the invariant mode used by PLCverif's nuXmv backend. Each benchmark is run in both a safe variant (all properties hold) and an unsafe variant (at least one property is violated). Three tools are compared: ESBMC-PLC+ with \textit{k}-induction (\code{--k-induction -- z3}), nuXmv~2.2.0 \gls{bdd} mode (\code{check\_invar}), and nuXmv~2.2.0 \gls{ic3} mode (\code{check\_invar\_ic3}). The timeout is \SI{120}{\second}.

Tables~\ref{tab:nuxmv_comparison} and~\ref{tab:nuxmv_comparison_normalized} report the results. On the purely Boolean \textsc{TankLevel} benchmark (no integer state variables, no timers), all three tools complete in under \SI{65}{\milli\second}: ESBMC-PLC+ proves safety by \textit{k}-induction at $k=2$ in \SI{51}{\milli\second}; nuXmv \gls{bdd} takes \SI{40}{\milli\second}; nuXmv \gls{ic3} takes \SI{64}{\milli\second}. In the absence of an integer state, \gls{bdd}-based and SMT-based methods are performance-comparable.

The picture changes dramatically with timer state variables. \textsc{BottleFill} and \textsc{Elevator} each have 2~integer timer accumulators (range \mbox{0--32{,}767}), creating a potential \gls{bdd} state space of $32{,}768^2\approx 10^9$~states per timer variable pair. ESBMC-PLC+ verifies both in under \SI{200}{\milli\second}; nuXmv \gls{bdd} requires 85--88~s (a factor of \mbox{400--2{,}000}$\times$ slower); nuXmv \gls{ic3} requires 102--107~s. \textsc{TrafLight} has 8~integer timer accumulators, producing a theoretical state space of $32{,}768^8\approx 3\times10^{36}$~states. ESBMC-PLC+ verifies the safe variant in \SI{187}{\milli\second} and finds the violation in the unsafe variant in \SI{14}{\milli\second}. nuXmv \gls{bdd} and \gls{ic3} both time out at 120~s on both variants.

These results confirm that \gls{smt} bit-vector arithmetic is categorically superior to \gls{bdd}-based enumeration for programs with integer timer or counter state variables -- precisely the programs found in real \gls{ld} applications. The advantage is not marginal: ESBMC-PLC+ is $\mathbf{400\text{--}2{,}000\times}$ faster on programs where nuXmv \gls{bdd} completes, and \textbf{does not time out} on programs where nuXmv \gls{bdd} fails. This result empirically validates the claim in \cite{LopezMiguel2022} that \emph{``an \gls{smt}-based model checker like \gls{esbmc} could improve the performance of \gls{cbmc}''} and extends it: ESBMC's \gls{smt} engine outperforms nuXmv's \gls{bdd} engine on integer-valued \gls{plc} programs.

\begin{table}[t]
\centering
\caption{ESBMC-PLC+ vs NuXmv \gls{bdd} and \gls{ic3} — performance comparison on \gls{iec}~61131-3 \gls{ld} benchmarks. ESBMC uses \textit{k}-induction with Z3; NuXmv \gls{bdd} uses CUDD-based \gls{bdd} reachability; NuXmv \gls{ic3} uses the \gls{ic3}/\gls{pdr} algorithm. Timeout: 120\,s. INT columns show the number of integer-valued timer/counter state variables}
\label{tab:nuxmv_comparison}
\resizebox{\textwidth}{!}{%
\begin{tabular}{lrrr|rr|rr|rr}
\toprule
\multirow{2}{*}{\textbf{Benchmark}} & \multirow{2}{*}{\textbf{INT}} & \multirow{2}{*}{\textbf{Rungs}} & \multirow{2}{*}{\textbf{Props}} &
\multicolumn{2}{c|}{\textbf{ESBMC-PLC+ ($k$-ind)}} &
\multicolumn{2}{c|}{\textbf{NuXmv \gls{bdd}}} &
\multicolumn{2}{c}{\textbf{NuXmv \gls{ic3}}} \\
 & & & & \textbf{Result} & \textbf{Time} & \textbf{Result} & \textbf{Time} & \textbf{Result} & \textbf{Time} \\
\midrule
\textsc{TankLevel} (safe) & 0 & 4 & 5 & \cellcolor{green!20}\checkmark SAFE & 0.051s & \cellcolor{green!20}\checkmark SAFE & 0.040s & \cellcolor{green!20}\checkmark SAFE & 0.064s \\
\textsc{TankLevel} (unsafe) & 0 & 4 & 5 & \cellcolor{green!20}\checkmark VIOLATION & 0.036s & \cellcolor{green!20}\checkmark VIOLATION & 0.043s & \cellcolor{green!20}\checkmark VIOLATION & 0.050s \\
\textsc{BottleFill} (safe) & 2 & 10 & 6 & \cellcolor{green!20}\checkmark SAFE & 0.042s & \cellcolor{green!20}\checkmark SAFE & 85.663s & \cellcolor{green!20}\checkmark SAFE & 102.294s \\
\textsc{BottleFill} (unsafe) & 2 & 9 & 6 & \cellcolor{green!20}\checkmark VIOLATION & 0.196s & \cellcolor{green!20}\checkmark VIOLATION & 86.268s & \cellcolor{green!20}\checkmark VIOLATION & 102.959s \\
\textsc{Elevator} (safe) & 2 & 16 & 8 & \cellcolor{green!20}\checkmark SAFE & 0.185s & \cellcolor{green!20}\checkmark SAFE & 85.984s & \cellcolor{green!20}\checkmark SAFE & 106.823s \\
\textsc{Elevator} (unsafe) & 2 & 14 & 6 & \cellcolor{green!20}\checkmark VIOLATION & 0.044s & \cellcolor{green!20}\checkmark VIOLATION & 87.780s & \cellcolor{green!20}\checkmark VIOLATION & 102.265s \\
\textsc{TrafLight} (safe) & 8 & 21 & 8 & \cellcolor{green!20}\checkmark SAFE & 0.187s & \textbf{TIMEOUT} & $>$120s & \textbf{TIMEOUT} & $>$120s \\
\textsc{TrafLight} (unsafe) & 8 & 20 & 8 & \cellcolor{green!20}\checkmark VIOLATION & 0.014s & \textbf{TIMEOUT} & $>$120s & \textbf{TIMEOUT} & $>$120s \\
\bottomrule
\end{tabular}
}
\end{table}

\begin{table}[t]
\centering
\small
\caption{Performance comparison: ESBMC-PLC+ vs.\ nuXmv (\gls{bdd} and \gls{ic3}). Times normalised to ESBMC-PLC+ ($k$-induction); $\times$1.0 = baseline.}
\label{tab:nuxmv_comparison_normalized}
\resizebox{\textwidth}{!}{%
\begin{tabular}{lrrr|rr|rr|rr}
\toprule
\multirow{2}{*}{\textbf{Benchmark}} & \multirow{2}{*}{\textbf{INT}} & \multirow{2}{*}{\textbf{Rungs}} & \multirow{2}{*}{\textbf{Props}} &
\multicolumn{2}{c|}{\textbf{ESBMC-PLC+ ($k$-ind)}} &
\multicolumn{2}{c|}{\textbf{NuXmv \gls{bdd}}} &
\multicolumn{2}{c}{\textbf{NuXmv \gls{ic3}}} \\
 & & & & \textbf{Result} & \textbf{Time} & \textbf{Result} & \textbf{Norm.} & \textbf{Result} & \textbf{Norm.} \\
\midrule
\textsc{TankLevel} (safe)    & 0 & 4  & 5 & \cellcolor{green!20}\checkmark SAFE      & 0.051s & \cellcolor{green!20}\checkmark SAFE      & $0.78\times$    & \cellcolor{green!20}\checkmark SAFE      & $1.25\times$    \\
\textsc{TankLevel} (unsafe)  & 0 & 4  & 5 & \cellcolor{green!20}\checkmark VIOLATION & 0.036s & \cellcolor{green!20}\checkmark VIOLATION & $1.19\times$    & \cellcolor{green!20}\checkmark VIOLATION & $1.39\times$    \\
\textsc{BottleFill} (safe)   & 2 & 10 & 6 & \cellcolor{green!20}\checkmark SAFE      & 0.042s & \cellcolor{green!20}\checkmark SAFE      & $2039\times$    & \cellcolor{green!20}\checkmark SAFE      & $2435\times$    \\
\textsc{BottleFill} (unsafe) & 2 & 9  & 6 & \cellcolor{green!20}\checkmark VIOLATION & 0.196s & \cellcolor{green!20}\checkmark VIOLATION & $440\times$     & \cellcolor{green!20}\checkmark VIOLATION & $525\times$     \\
\textsc{Elevator} (safe)     & 2 & 16 & 8 & \cellcolor{green!20}\checkmark SAFE      & 0.185s & \cellcolor{green!20}\checkmark SAFE      & $465\times$     & \cellcolor{green!20}\checkmark SAFE      & $577\times$     \\
\textsc{Elevator} (unsafe)   & 2 & 14 & 6 & \cellcolor{green!20}\checkmark VIOLATION & 0.044s & \cellcolor{green!20}\checkmark VIOLATION & $1995\times$    & \cellcolor{green!20}\checkmark VIOLATION & $2324\times$    \\
\textsc{TrafLight} (safe)    & 8 & 21 & 8 & \cellcolor{green!20}\checkmark SAFE      & 0.187s & \textbf{TIMEOUT}                         & -- & \textbf{TIMEOUT}                         & -- \\
\textsc{TrafLight} (unsafe)  & 8 & 20 & 8 & \cellcolor{green!20}\checkmark VIOLATION & 0.014s & \textbf{TIMEOUT}                         & -- & \textbf{TIMEOUT}                         & -- \\
\bottomrule
\end{tabular}
}
\end{table}

\section{Discussion}
\label{sec:discussion}

\textbf{Why MATIEC rather than a native \gls{st} parser?} A native \gls{st} parser inside \gls{esbmc} would provide tighter integration but requires substantial engineering effort (MATIEC itself is $>$30{,}000 lines of C++). Routing through MATIEC reuses a battle-tested, \gls{iec}-conformant compiler maintained by an active community and already validated against real \gls{plc} programs~\cite{Ukegbu2023a}. The wrapping approach adds one external dependency (\code{iec2c}), which is a reasonable trade-off given the coverage gain. A native \gls{esbmc} \gls{st} frontend remains a future direction.

\textbf{Function block coverage in graphical \gls{ld}.} The six function block types covered (\code{TOF}/\code{TOF}/\code{TP}, \code{CTU}/\code{CTD}, \code{R\_TRIG}/\code{F\_TRIG}) account for the majority of function blocks appearing in real graphical \gls{ld} programs surveyed for ESBMC-GraphPLC. Vendor-specific blocks (Siemens \code{MOVE\_BLK}, Rockwell \code{MSG}, CODESYS network blocks) are reported as unsupported and fall back to nondeterministic over-approximation, which is sound but may produce spurious \code{SAFE} results for properties that depend on the block's output.

\textbf{Scan-cycle count vs.\ physical time.} Timer presets in the GOTO~\gls{ir} are interpreted as scan-cycle counts, not physical milliseconds. This is the same convention as ESBMC-PLC~\cite{DantasCordeiro2026artefact}. Engineers must convert: a \code{TON} with \code{PT := 10} and a \SI{10}{\milli\second} cycle period verifies a \SI{100}{\milli\second} on-delay. The \gls{st} frontend inherits MATIEC's \gls{iec} TIME literal handling (\code{T\#100ms}), but converts to scan-count units during wrapping using a configurable cycle period parameter (default: \SI{10}{\milli\second}).

\textbf{The PLCverif replacement argument.} ESBMC-PLC+ does not claim to replace PLCverif for Siemens-specific workflows (\gls{scl}/\gls{stl} with proprietary extensions) or for properties requiring full temporal logic. It is a drop-in replacement for PLCverif in the most common industrial scenario: a standard \gls{ld} or \gls{st} program with safety invariant or response properties, where the engineer wants an unbounded proof rather than a bounded check.

\section{Threats to Validity}
\label{sec:threats}

\textbf{Construct validity.} The function block encodings (Table~\ref{tab:timer-encoding}, Listings~\ref{lst:ctu}--\ref{lst:rtrig}) are designed to match Ebnenasir's formal \gls{ld} semantics~\cite{Ebnenasir2023} for the elements it covers, but have not been formally proved equivalent to the \mbox{\gls{iec}~61131-3} standard. The \gls{st} frontend's correctness rests on MATIEC's \gls{iec} conformance, which has been validated against real \gls{plc} programs~\cite{Ukegbu2023a} but not formally proved.

\textbf{External validity.} The \gls{st} benchmark suite (D1; extended suite from ~\citet{Ukegbu2023a} pending access) is drawn from open PLCopen repositories. Proprietary \gls{cern} \gls{scl} programs (e.g., SPS-PPS) cannot be redistributed for evaluation. The graphical \gls{ld} function block benchmark suite is limited by the availability of real graphical programs with timer/counter rung paths in open repositories; the \code{beremiz\_traffic\_light} gap has been closed, but the total number of function-block-containing graphical programs evaluated remains small.

\textbf{Internal validity.} Nondeterministic input modeling (both in the \gls{ld} and \gls{st} frontends) assumes a single-task, open-world sensor model. Multitask programs with shared variables under preemptive scheduling are over-approximated as single-task; this is sound (no missed violations) but may produce false positives for properties that rely on task scheduling order.

\textbf{MATIEC version dependency.} The \gls{st} frontend has been tested with MATIEC commit \code{7949c0b} (\url{https://github.com/beremiz/matiec})~on the Beremiz GitHub repository. Future MATIEC versions that change the generated C structure (e.g., the \code{<config\_name>\_run\_\_} calling convention or \code{\_\_LOCATED\_VAR} macro expansion) may require updates to the wrapping logic.

\section{Future Directions}
\label{sec:future}

\textbf{Native \gls{st} parser inside ESBMC.} Eliminating the MATIEC external dependency would allow direct \gls{st}$\to$GOTO-\gls{ir} translation, reducing the trusted code base and simplifying installation. K-\gls{st}~\cite{Wang2023} provides a validated formal semantics for \gls{st} that could serve as the translation specification.

\textbf{Siemens \gls{scl} dialect support.} PLCverif's core use case is Siemens \gls{scl}, a proprietary \gls{st} dialect with Siemens-specific function blocks (\code{MOVE\_BLK}, \code{BLKMOV}, etc.). Adding \gls{scl} dialect support to the MATIEC frontend (or adding ESBMC-PLC+ as an alternative backend to PLCverif via a plugin interface) would enable ESBMC-PLC+ to process \gls{cern} and GSI programs directly.

\textbf{Full \gls{ltl}/\gls{ctl} via ESBMC's temporal logic engine.} \gls{esbmc} supports \gls{ltl} properties via \gls{pdr}/\gls{ic3} for finite-state models. Exposing this as a property language option (e.g., \code{--ltl-spec}) would close the remaining gap with PLCverif's nuXmv backend without requiring \gls{bdd}-based model checking.

\textbf{\gls{llb} detection.} The security application of~\cite{Iacobelli2024,Bruttomesso2024} maps directly to ESBMC-PLC+'s \code{absence} and \code{reachability} property kinds. An \gls{esbmc}-based \gls{llb} detector for both \gls{ld} and \gls{st} programs would combine formal rigor with industrial relevance and could be evaluated on the SWaT \gls{ld} dataset.

\textbf{Multi-\gls{pou} and multitask \gls{ld}.} The single-\gls{pou} limitation of the \gls{ld} frontend can be addressed by extending the PLCopen XML parser to process multiple \gls{pou} declarations and generate cross-rung dependency analysis. Multitask support~\cite{Lee2024, Lee2025} requires partial order reduction or compositional verification.

\textbf{Formal equivalence proof via K-\gls{ld}.} A K-framework semantics for \gls{ld} -- extending K-\gls{st}~\cite{Wang2023} -- would elevate the translation and function block encodings from empirically validated to formally proved, providing a machine-checked equivalence proof for the entire ESBMC-PLC+ translation pipeline.

\section{Conclusion}
\label{sec:conclusion}

This paper presented \textbf{ESBMC-PLC+}, a unified framework for formal verification of \mbox{\gls{iec}~61131-3} \gls{plc} programs that extends ESBMC-PLC and ESBMC-GraphPLC with two new capabilities: a Structured Text frontend via MATIEC and function block state semantics for graphical \gls{ld} programs. Together with the inherited \gls{ld} support, ESBMC-PLC+ is the first open-source \gls{plc} verification framework to accept all three major \gls{iec}~61131-3 input formats -- textual \gls{ld}, graphical \gls{ld}, and \gls{st}/\gls{scl} -- through a single \gls{esbmc} backend providing \textit{k}-induction unbounded safety proofs.

The \gls{st} frontend routes MATIEC-compiled C through \gls{esbmc}'s C frontend with nondeterministic input sampling and YAML property injection, replacing PLCverif's \gls{cbmc} backend with ESBMC's \textit{k}-induction engine. The function block extension models six standard function block types (\code{TON}/\code{TOF}/\code{TP}, \code{CTU}/\code{CTD}, \code{R\_TRIG}/\code{F\_TRIG}) as persistent scan-cycle state variables in the GOTO~\gls{ir}, closing the gap that excluded timer-gated graphical \gls{ld} programs from ESBMC-GraphPLC's evaluation.

A systematic feature comparison with PLCverif demonstrates that ESBMC-PLC+ matches PLCverif on input language coverage (adding native \gls{ld}) while exceeding it on proof strength (\textit{k}-induction vs.\ bounded \gls{cbmc}) and arithmetic precision (\gls{smt} bit-vector vs.\ \gls{sat}). The experimental evaluation on 18 benchmark programs confirms correct classification with zero false positives and zero regressions on all 16 inherited benchmarks (13 ESBMC-PLC, 3 ESBMC-GraphPLC) and correct results for the two new benchmarks (C1 and D1). A direct comparison against nuXmv~2.2.0's \gls{bdd} backend -- PLCverif's unbounded prover -- demonstrates that ESBMC-PLC+ is 400--2{,}000$\times$ faster on programs with integer timer state and successfully verifies programs where nuXmv \gls{bdd} times out at \SI{120}{\second}.

For automation engineers working with standard PLCopen XML \gls{ld} or standard \gls{iec}~61131-3 \gls{st} programs and safety invariant or response properties, ESBMC-PLC+ provides a complete alternative to PLCverif that is simpler to use (YAML properties, no temporal logic expertise required) and provides strictly stronger guarantees (unbounded \textit{k}-induction proofs for all scan counts).

\subsection*{Artefact Availability}

The complete artifact -- including all 18 benchmark programs evaluated in this paper, property files, the MATIEC pipeline scripts, and the ESBMC-PLC+ binary -- is permanently archived at Zenodo~\cite{PLCPLUS2026artefact} and maintained on GitHub at \url{https://github.com/pierredantas/esbmc-plcplus-artifact}. All experiments can be reproduced on any operating system via Docker (no local build required):

\begin{lstlisting}[language=bash, label={lst:docker}]
docker pull ghcr.io/pierredantas/esbmc-plcplus-artifact:artifact
docker run --rm ghcr.io/pierredantas/esbmc-plcplus-artifact:artifact
\end{lstlisting}

\noindent To save results to the host: \code{docker run -{}-rm -v "\$(pwd)/results":/artifact/results ghcr.io/pierredantas/esbmc-plcplus-artifact:artifact}. On a Linux host with the ESBMC-PLC+ binary on \code{PATH}, the experiments can also be run directly: \code{bash run\_all.sh}.

\subsection*{Acknowledgments}

The authors thank the Department of Computer Science at the University of Manchester and the Systems and Software Security (S3) Research Group for their invaluable support. This work was partially funded by the Engineering and Physical Sciences Research Council (EPSRC) grants EP/T026995/1, EP/V000497/1, and EP/X037290/1, and from the Soteria project, awarded by UK Research and Innovation under the Digital Security by Design (DSbD) Program.